\documentclass[9pt,twocolumn,twoside,lineno]{pnas-new}
\usepackage{subfigure}
\usepackage{import}
\newcommand {\E} {\mathbb{E}}

\newcommand{\pr}{\mathbb{P}}

\newcommand{\tfone}{T^f_1}
\newcommand{\tftwo}{T^f_2}
\newcommand{\twone}{T^w_1}
\newcommand{\twtwo}{T^w_2}
\newcommand{\mufoneone}{\mu^f_{11}}
\newcommand{\mufonetwo}{\mu^f_{12}}
\newcommand{\muftwoone}{\mu^f_{21}}
\newcommand{\muftwotwo}{\mu^f_{22}}
\newcommand{\muoneone}{\mu_{11}}
\newcommand{\muonetwo}{\mu_{12}}
\newcommand{\mutwoone}{\mu_{21}}
\newcommand{\mutwotwo}{\mu_{22}}
\newcommand{\muwoneone}{\mu^w_{11}}
\newcommand{\muwonetwo}{\mu^w_{12}}
\newcommand{\muwtwoone}{\mu^w_{21}}
\newcommand{\muwtwotwo}{\mu^w_{22}}

\newcommand{\qfone}{q^f_1}
\newcommand{\qftwo}{q^f_2}
\newcommand{\qwone}{q^w_1}
\newcommand{\qwtwo}{q^w_2}
\newcommand{\gammaoneQ}{\gamma_1(q^f_1,q^f_2, q^w_1,q^w_2 )}
\newcommand{\GammafoneQ}{\Gamma^f_1(q^f_1,q^f_2, q^w_1,q^w_2)}
\newcommand{\GammaftwoQ}{\Gamma^f_2(q^f_1,q^f_2, q^w_1,q^w_2)}
\newcommand{\GammawoneQ}{\Gamma^w_1(q^f_1,q^f_2, q^w_1,q^w_2)}
\newcommand{\GammawtwoQ}{\Gamma^w_2(q^f_1,q^f_2, q^w_1,q^w_2)}

\templatetype{pnasresearcharticle} % Choose template 

\title{Spreading Processes with Mutations over Multi-layer Networks}

\author[a]{Mansi Sood}
\author[b]{Anirudh Sridhar} 
\author[c]{Rashad Eletreby}
\author[d]{Chai Wah Wu}
\author[e]{Simon A. Levin}
\author[b]{H. Vincent Poor}
\author[a]{Osman Yagan}

\affil[a]{Department of Electrical and Computer Engineering, Carnegie Mellon University, Pittsburgh, PA 15213 USA}
\affil[b]{Department of Electrical Engineering, Princeton University, Princeton, NJ 08544 USA}
\affil[c]{Rocket Travel, Inc, Chicago, IL 60661 USA}
\affil[d]{Thomas J. Watson Research Center, IBM, Yorktown Heights, NY 10598 USA}
\affil[e]{Department of Ecology and Evolutionary Biology, Princeton University, Princeton, NJ 08544 USA}
\leadauthor{Sood}

\correspondingauthor{\textsuperscript{2}To whom correspondence should be addressed. E-mail: \texttt{msood@andrew.cmu.edu}}


\begin{abstract}
A key scientific challenge during the outbreak of novel infectious diseases is to predict how the course of the epidemic changes under different countermeasures that limit interaction in the population. Most epidemiological models do not consider the role of mutations and heterogeneity in the type of contact events. However, pathogens have the capacity to \emph{mutate} in response to changing environments, especially caused by the increase in population immunity to existing strains and the emergence of new pathogen \emph{strains} poses a continued threat to public health. Further, in light of differing transmission risks in different {congregate settings} (e.g., schools and offices), different mitigation strategies may need to be adopted to control the spread of infection. We analyze a \emph{multi-layer multi-strain} model by simultaneously accounting for i) pathways for mutations in the pathogen leading to the emergence of new pathogen {strains}, and ii) differing transmission risks in different congregate settings, modeled as \emph{network-layers}. Assuming \emph{complete cross-immunity} among strains, namely, recovery from any infection prevents infection with any other (an assumption that will need to be relaxed to deal with COVID-19 or influenza), we derive the key epidemiological parameters for the proposed multi-layer multi-strain framework. We demonstrate that reductions to existing network-based models that discount heterogeneity in either the strain or the network layers can lead to incorrect predictions for the course of the outbreak. In addition, our results highlight that the impact of imposing/lifting mitigation measures concerning different contact network layers (e.g., school closures or work-from-home policies) should be evaluated in connection with their effect on the likelihood of the emergence of new pathogen strains.
\end{abstract}
\authorcontributions{Author contributions: M.S., H.V.P., and O.Y. designed research; M.S., A.S., C.W.W., and O.Y. performed research; M.S., A.S., C.W.W., S.A.L, H.V.P., and O.Y. contributed new reagents/analytic tools; M.S., A,S., R.E., S.A.L., H.V.P. and O.Y. analyzed data; and M.S., A.S., R.E., C.W.W., S.A.L., H.V.P., and O.Y. wrote the paper.}
\begin{document}
	\maketitle
	\thispagestyle{firststyle}
	\ifthenelse{\boolean{shortarticle}}{\ifthenelse{\boolean{singlecolumn}}{\abscontentformatted}{\abscontent}}{}

		\section*{Introduction}\dropcap{T}he recent outbreak of the COVID-19 pandemic, fuelled by the novel coronavirus SARS-CoV-2 led to a devastating loss of human life and upended livelihoods worldwide \cite{who-article}. The highly transmissible, virulent, and rapidly mutating nature of the SARS-CoV-2 coronavirus \cite{harvey2021sars} led to an unprecedented burden on critical healthcare infrastructure. The emergence of new \emph{strains} of the pathogen as a result of \emph{mutations} poses a continued risk to public health \cite{islam2022new, tregoning2021variants}. Moreover, when a new {strain} is introduced to a host population, pharmaceutical interventions often take time to be developed, tested, and made widely accessible \cite{burgos2021race,beyrer2021human}. In the absence of widespread access to treatment and vaccines, policymakers are faced with the challenging problem of taming the outbreak with \emph{nonpharmaceutical interventions} (NPIs) that encourage physical distancing in the host population to suppress the growth rate of new infections \cite{matrajt2020evaluating, eubank2020commentary,morris2021optimal}. However, the ensuing socio-economic burden \cite{mandel2020economic,arndt2020covid} of NPIs, such as lockdowns, makes it necessary to understand how imposing restrictions in different social settings (e.g., schools, offices, etc.) alter the course of the epidemic outbreak. %From the standpoint of epidemiological modeling, a key challenge is to develop and analyze models that simultaneously account for patterns of interaction in the host population and pathways of evolutionary adaptions in the pathogen.

Epidemiological models that analyze the {speed} and {scale} of the spread of infection can be broadly classified under two approaches. The first approach assumes {\em homogeneous mixing}, i.e., the population is well-mixed, and an infected individual is equally likely to infect any individual in the population regardless of location and social interactions \cite{brauer2008compartmental,anderson1992infectious}. The second is a {\em network-based} approach that explicitly models the contact patterns among individuals in the population and the probability of transmission through any given contact \cite{newman2002spread,kenah2007second,Salath22020}. Structural properties of the contact network such as {\em heterogeneity} in type of contacts \cite{sun2021transmission}, {clustering} (e.g., presence of tightly connected {communities}) \cite{zhuang2017clustering}, {centrality} (e.g., presence of super-spreaders) \cite{s2021superspreading,zeng2021identifying} and {degree-degree correlations} \cite{so2021topological} are known to have profound implications for disease spread and its control \cite{hebert2010propagation, thurner2020network}. %to study the impact of countermeasures, we focus on a network-based epidemiological model.
To understand the impact of NPIs that lead to reduction in physical contacts, network-based epidemiological models have been employed widely in the context of infectious diseases, including COVID-19 \cite{ marathe_network, bongiorno2022multi, aleta2020modelling}. %[more multi-layer info]

In addition to the contact structure within the host population, the course of an infectious disease is critically tied to {\em evolutionary adaptations} or {\em mutations} in the pathogen. There is growing evidence for the zoonotic origin of disease outbreaks, including COVID-19, SARS, and H1N1 influenza, as a result of cross-species transmission and subsequent evolutionary adaptations \cite{ne:alexday,ye2020zoonotic,cui2019origin, latinne2020origin,parrish2008cross}. When pathogens enter a new species, they are often poorly adapted to the physiological environment in the new hosts and undergo evolutionary mutations to adapt to the new hosts \cite{ne:alexday}. The resulting {\em variants} or {\em strains} of the pathogen have varying risks of transmission, commonly measured through the reproduction number or $R_0$, which quantifies the mean number of secondary infections triggered by an infected individual \cite{AndersonMay,Vespignanir0}. Moreover, even when a sizeable fraction of the population gains immunity through vaccination or natural infection, the emergence of new variants that can evade the acquired immunity poses a continued threat to public health \cite{islam2022new,tregoning2021variants}. A growing body of work \cite{ne:alexday,ne:pnas,girvan2002simple,andreasen1997dynamics,kryazhimskiy2007state,phylodynamic-science.1090727,phylo-o2011contact, phylo-holmes2009discovering-network, phylo-shiino2012phylodynamic, marquioni2021modeling,zhang2021epidemic,yule1925ii} has highlighted the need for developing \emph{multi-strain} epidemiological models that account for evolutionary adaptations in the pathogen. For instance, there is a vast literature on \emph{phylodynamics} \cite{phylodynamic-science.1090727,phylo-o2011contact, phylo-holmes2009discovering-network, phylo-shiino2012phylodynamic} which examines how epidemiological and evolutionary processes interact to impact pathogen {phylogenies}. The past decade has also seen the development of \emph{network-based} models to identify risk factors for the emergence of pathogens in light of different contact patterns \cite{ne:alexday,ne:pnas,marquioni2021modeling,zhang2021epidemic}. Further, a recent study \cite{ne:pnas} demonstrated that models that do not consider evolutionary adaptations may lead to incorrect predictions about the probability of the emergence of an epidemic triggered by a mutating contagion. %and intra-host/inter-host pathogen evolution

Most existing network-based approaches that analyze the spread of mutating pathogens center on single-layered contact networks, where the {\em transmissibility}, i.e., the probability that an infective individual passes on the infection to a contact, depends on the type of \emph{strain} but \emph{not} on the nature of \emph{link/contact} over which the infection is transmitted \cite{ne:alexday,ne:pnas, zhang2021epidemic}. However, different congregate settings such as schools, hospitals, offices, and private social gatherings pose varied transmission risks \cite{ajelli2014role}. Recently, \emph{multi-layer} networks have been used to model human contact networks \cite{marathe_network, bongiorno2022multi, aleta2020modelling, Vespignanir0}, where each layer represents a different social setting in which an individual participates. While \emph{multi-layer} contact networks \cite{ne:conjoining,haccett-percolation-multiplex,Bianconi_2017,yong_tnse, mann-arbitrary-clustering,sahneh2013generalized,zeng2021identifying,marceau2011modeling,radicchi2015percolation,PhysRevEAzimi}  and {multi-strain} contagions \cite{ne:alexday,ne:pnas,girvan2002simple,andreasen1997dynamics,kryazhimskiy2007state, zhang2021epidemic} have been extensively studied in separate contexts, there has been a dearth of analysis on {simultaneously} accounting for the multi-strain network structure and multi-strain spreading. 

In this paper, we build upon the mathematical theory for the \emph{multi-strain} model proposed in \cite{ne:alexday, ne:pnas} to  account for the {\em multi-layer} structure typical to human contact networks, where different network layers correspond to different social settings in which individuals congregate. Specifically, we assume that the {\em transmissibility} depends not only on the type of strain carried by an infective individual but also on the social setting (modeled through a network layer) in which they meet their neighbors. The proposed framework allows studying how NPIs, such as lockdowns in different network layers, impact the course of the spread of a contagion.

While the bulk of our discussion is on mutating contagions in the context of infectious diseases, our results also hold promise for applications in modeling social contagions, e.g., news items circulating in social networks \cite{ne:pnas, dawkins2016selfish}. Similar to different strains of a pathogen arising through mutations, different versions of the information are created as the content is altered on social media platforms \cite{Adamic2016}. The resulting \emph{variants} of the information may have varying propensities to be circulated in the social network. Moreover, with the burst of social media platforms, potential applications of our multi-layer analysis of mutating contagions are in analyzing the multi-platform spread of misinformation where the information gets altered across different platforms.

\section*{Model}
		\label{sec:Model}
	%	\onecolumn
	\subsection*{Contact Network}
 We consider a population of size $n$ with members in the set $\mathcal{N}=\{1,\ldots, n\}$. Patterns of interaction in the host population are encoded in the contact network where an edge is drawn between two nodes if they can come in contact and potentially transmit the infection. To account for variability in transmission risks associated with different social settings (e.g., neighborhood, school, workplace), we consider a \emph{multi-layer} contact network \cite{ne:conjoining}, where each network layer corresponds to a unique social setting. For simplicity, we focus on the case where each individual can participate \emph{independently} in two network layers denoted by $\mathbb{F}$ and $\mathbb{W}$ respectively. %{\color{red} (The following sentence needs a rewrite)} 
 %These network layers represent interactions individuals make with those they come in contact with in different congregate settings. 
 For instance, the network layer $\mathbb{F}$ can be used to model the spread of infection between friends residing in the same neighborhood, while the network layer $\mathbb{W}$ can model the spread of infections amongst individuals who congregate for work. 
		\begin{figure}[tbhp]
		\centering
		\includegraphics[scale=0.19]{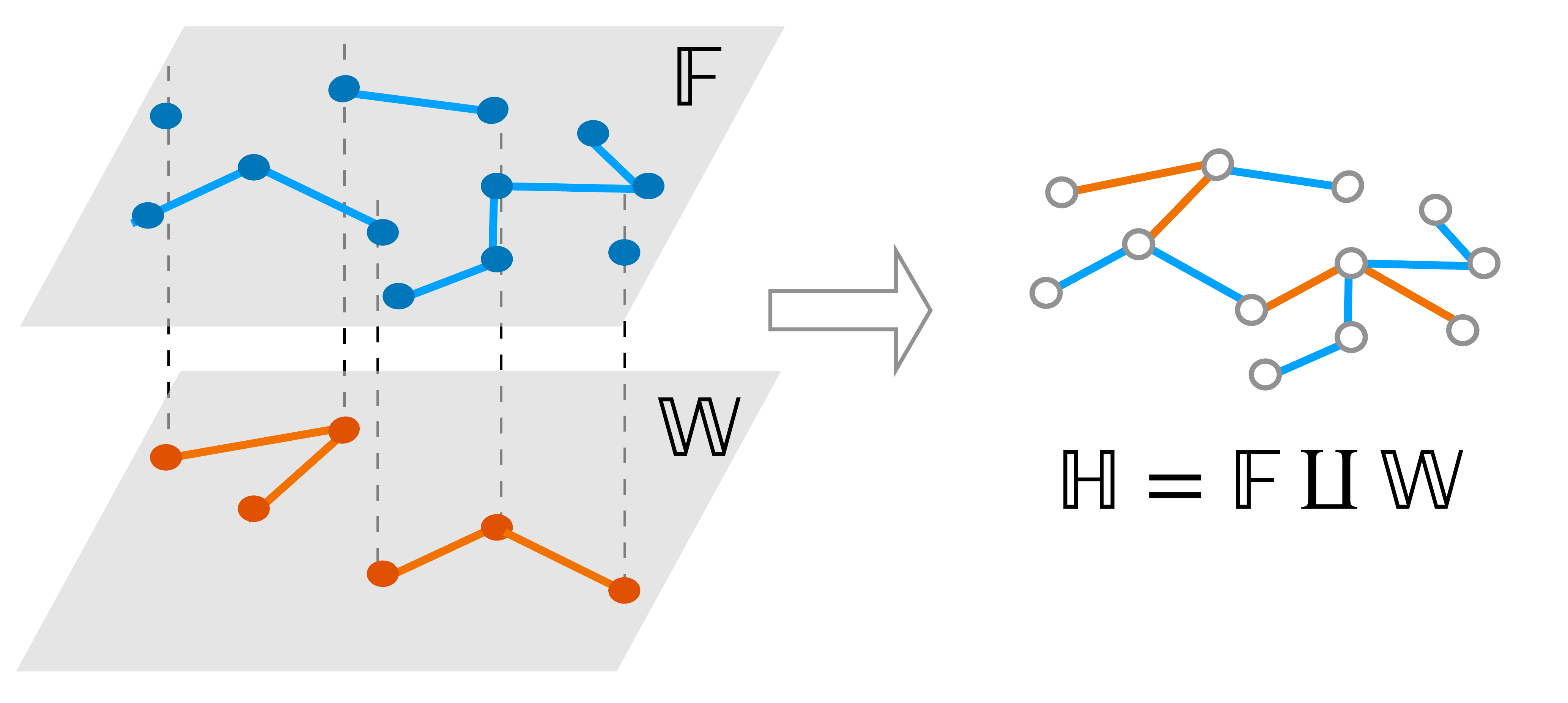}
		\caption{\sl {\bf Multi-layer network model:} An illustration of a two-layer contact network for modeling the spread of an infection over the friendship/neighborhood network $\mathbb{F}$ and work network~$\mathbb{W}$. The resultant contact network $\mathbb{H}=\mathbb{F} \amalg \mathbb{W}$. Neighboring nodes in $\mathbb{H}$ can transmit infections to their neighbors either through links in the $\mathbb{F}$ network (i.e., through type-$f$ links) or $\mathbb{W}$ network (through type-$w$ links).} 
		\label{fig:ne:networkmodel}
	\end{figure} %{\color{red} (I think we should mention `in network layer a`)}
In order to model participation in each network layer, we first independently label each node as \emph{non-participating} in network layer-$a$ with probability $\alpha_a$ and \emph{participating} in network layer-$a$ with probability $1-\alpha_a$, where $0\leq \alpha_a \leq 1$, and where $a \in \{f,w\}$. Next, for each node that {participates} in network layer-$a$, the number of its neighbors in layer-$a$ is drawn from a degree distribution, denoted by $\{\tilde{p}^a_k, k=0,1,\dots,n\}$, where $a \in \{f,w\}$. Under this formulation, the degree of a node in layer-$a$, denoted by $\{p^a_k, \: k=0,1,\ldots\}$, with $a \in \{f,w\}$, is given by
	\begin{align}
		p^a_k = (1-\alpha_a)\tilde{p}^a_k + \alpha_a \mathbf{1} \{k=0\},~~~ k=0,1,\dots, \label{eq:deg-first}
	\end{align}
where $\mathbf{1}\{\}$ denotes the indicator random variable, admitting the value one when $k=0$ and zero when $k
\geq 1$. We generate both layers independently
	according to the {\em configuration} model
	\cite{Bollobas,MolloyReed} with the degree distribution given through \eqref{eq:deg-first}. For notational simplicity, we say that edges in network $\mathbb{F}$ (resp., $\mathbb{H}$) are of type-$f$ (resp., type-$w$). The multi-layer contact network, denoted as  $\mathbb{H}$, is constructed by taking the disjoint union $(\amalg)$ of network layers $\mathbb{W}$ and
	$\mathbb{F}$ (Figure~\ref{fig:ne:networkmodel}). We assume that the network $\mathbb{H}$ is static and focus on the emergent spreading behavior in the limit of infinite population size $(n \rightarrow \infty)$.

	\subsection*{Spreading Process}
		%{\color{red} (I understand that we are targeting generalization by making mutation probabilities different for different layers, but perhaps we need to make a case for this generalization. One concern I have is that reviewers might ask why do mutations change across layers, isn't mutation an inherent property of the pathogen itself independent from the layers? So we probably need to give an example or generally make a case why different mutations are needed.)} 
  We adopt a \emph{multi-strain} spreading process \cite{ne:alexday} to the \emph{multi-layer} network setting as follows. For each layer, the evolutionary adaptations in the pathogen are modeled by corresponding mutation matrices. Let $m$ denote the number of pathogen strains co-existing in a population. 
  For network layer $\mathbb{F}$ (resp., $\mathbb{W}$), the mutation matrix, denoted by $\pmb{\mu}^f$ (resp., $\pmb{\mu}^w$) is a $m \times m$ matrix. The entry $\mu^f_{ij} $ (resp., $\mu^w_{ij}$) denotes the probability that strain-$i$ mutates to strain-$j$ within a host who got infected through a type-$f$ (resp., type-$w$) link, with $\sum_{j} \mu^f_{ij}=1$ (resp., $\sum_{j}\mu^w_{ij}=1$). %Motivated by settings where the epidemiological and evolutionary processes occur at a similar time-scale and mutations of the pathogen occur within the host, we assume that the mutation probability depends on the type of incoming link that was used to infect a node and not on the type of links through which the infected node subsequently infects their neighbors. 
	Given that an individual carrying strain-$i$ makes an infectious contact through a type-$f$ (resp., type-$w$) edge, the newly infected individual acquires strain $j$ with probability $\mu^f_{ij}$ (resp., $\mu^w_{ij}$). We note that for the context of infectious diseases where the epidemiological and evolutionary processes occur at a similar time-scale and mutations of the pathogen occur within the host, the mutation matrices do not depend on the network structure \cite{ne:alexday} and $\pmb{\mu}^w=\pmb{\mu}^f$ \footnote{In the context of information propagation, different strains \cite{ne:pnas} may correspond to different versions of the information. Therefore, to provide a more general contagion model, we let the mutation probabilities depend on the network layer.}.

\begin{figure}[tbhp]
		\centering
		\includegraphics[scale=0.17]{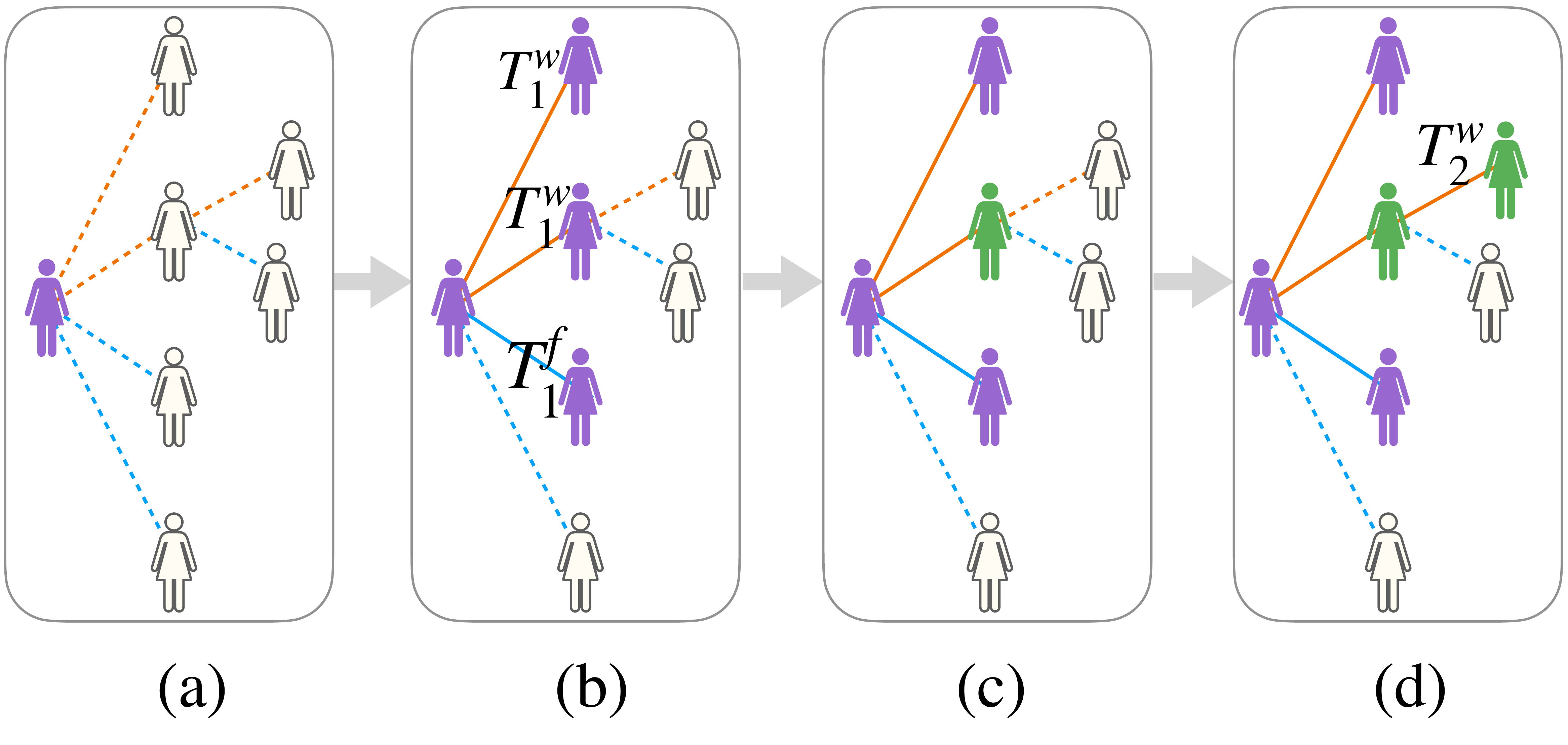}
		\caption{\sl {\bf Multi-strain transmission model:} An illustration of the multi-strain model with 2 strains on a contact network comprising 2 layers--  (a) An arbitrary chosen seed node acquires strain-1; (b) The seed node independently infects their susceptible neighbors connected through type-$f$ (resp., type-$w$) links with probability $T^f_1$ (resp., $T^w_1$); (c) After infection, the pathogen mutates to strain-2 within the hosts with probabilities given by mutation matrices $\pmb{\mu}^f$ and $\pmb{\mu}^w$; (d) The infected nodes in turn infect their neighbors with transmission probabilities governed by the strain that they are carrying (i.e, strain-1 or strain-2), and the type of edge  used to infect their neighbors (i.e., type-1 or type-2). The process terminates when no further infections are possible.} 
		\label{fig:ne:txmodel}
	\end{figure}

 In the succeeding discussion, we focus on the setting where two strains of the pathogen are dominant and assume $m=2$. We denote
	\begin{align}
		\pmb{\mu}^f= \left[   \begin{matrix}
			\mufoneone &  \mufonetwo\\ 
			\muftwoone &  \muftwotwo
		\end{matrix} \right], \nonumber  \qquad 
		\pmb{\mu}^w= \left[    \begin{matrix}
			\muwoneone &  \muwonetwo\\ 
			\muwtwoone &  \muwtwotwo
		\end{matrix} \right]. \nonumber 
	\end{align}
	We model the dependence of transmissibility on the type of links using $m \times m$  diagonal matrices $\pmb{T}^f$ (resp., $\pmb{T}^w$), with $[T^f_i]$ (resp., $[T^w_i]$) representing the transmissibility of strain-$i$ over a type-$f$ link (resp., type-$w$ link), for $i=1,\dots,m$. We have
	\begin{align}
		\pmb{T}^f= \left[   \begin{matrix}
			\tfone &  0\\ 
			0 &  \tftwo
		\end{matrix} \right], \nonumber  \qquad 
		\pmb{T}^w= \left[    \begin{matrix}
			\twone &  0\\ 
			0 &  \twtwo
		\end{matrix} \right]. \nonumber 
	\end{align}
	%\emph{Transmissibility:}
	We consider the following multi-strain spreading process on a multi-layer network (Figure~\ref{fig:ne:txmodel}) that accounts for pathogen transmission when epidemiological and evolutionary processes occur on a similar timescale and each new infection offers an opportunity for mutation \cite{ne:alexday}. The process starts when a randomly chosen seed node is infected with strain-1. We refer to such a seed node as the {\em initial infective} and the nodes that are subsequently infected as {\em later-generation infectives}. The seed node independently infects their susceptible neighbors connected through type-$f$ (resp., type-$w$) links with probability $T^f_1$ (resp., $T^w_1$). We assume that co-infection is not possible and after infection, the pathogen mutates to strain-$i$ within the hosts with probabilities given by mutation matrices $\pmb{\mu}^f$ and $\pmb{\mu}^w$. Further, in line with \cite{ne:alexday,ne:pnas}, we assume that once an individual becomes \emph{recovered} after being infected with either strain, then they can not be reinfected with \emph{any} strain. The infected nodes in turn infect their neighbors independently with transmission probabilities governed by the strain that they are carrying (i.e, strain-1 or strain-2), and the type of edge  used to infect their neighbors (i.e., type-$f$ or type-$w$). The process terminates when no further infections are possible.  Additional details regarding the Materials and Methods are presented in \emph{\color{blue}SI Appendix 1}.%
	
 We note that this paper is the first effort to develop a framework for the multiscale process discussed.  In it, we assume \emph{complete cross-immunity} between strains: recovery from any infection prevents infection with any other.  This is a good assumption for example for myxomatosis, but is not a good assumption for influenza or COVID, for which the emergence of new strains is driven by escape from population immunity.   The case of incomplete cross-immunity, which is an essential feature of the current pandemic,  therefore will be the subject of a follow-up paper.

	\section*{Results}
	\subsection*{Summary of key contributions}
We provide analytical results for characterizing epidemic outbreaks caused by mutating pathogens over multi-layer contact networks using tools from multi-type branching processes. In particular, we derive three key metrics to quantify the epidemic outbreak: i) the probability of emergence of an epidemic, ii) the expected fraction of individuals infected with each strain, and iii) the critical threshold of phase transition beyond which an epidemic outbreak occurs with a positive probability. Specifically,  the probability of emergence is defined as the probability that a randomly chosen infectious seed node leads to an epidemic, i.e., a positive fraction of nodes get infected in the limit of large network size. The epidemic threshold defines the {\em critical} point at which a phase transition occurs, leading to the possibility of an epidemic outbreak. In other words, the epidemic threshold defines a region in the parameter space in which the epidemic occurs with a positive probability while outside that region, the outbreak dies out after a finite number of transmissions. Finally, we derive the conditional mean of the fraction of individuals who get infected by each type of strain given that an epidemic outbreak has occurred.

We supplement our theoretical findings with analytical case studies and simulations for different patterns of interaction in the host population and different types of mutation patterns in the pathogens. The multi-layer multi-strain modeling framework allows for understanding trade-offs, such as the relative impact of countermeasures, including lock-downs that alter the network layers on the emergence of highly contagious strains. For cases where the spread of infection starts with a moderately transmissible strain, we study how imposing/lifting mitigation measures across different layers can alter the course of the epidemic by increasing the risk of mutation to a highly contagious strain. We derive the probability of mutation to a highly transmissible strain which in turn provides a lower bound on the probability of emergence. Through a case study for \emph{one-step irreversible} mutation patterns, our results highlight that reopening a new layer in the contact network may be considered low-risk based on the transmissibility of the current strain. Still, even a modest increase in infections caused by the additional layer can lead to an epidemic outbreak to occur with a much higher probability. Therefore, it is important to evaluate mitigation measures concerning different network layers in connection with their impact on the likelihood of the emergence of new pathogen strains.
%{\color{red} (I think the reviewer will say, ``Okay, given that it could be mild or dangerous, what's the conclusion here, or what should we do basically?")}

Next, we propose  transformations to simpler epidemiological models and unravel conditions under which we can reduce the multi-layer multi-strain model to simpler models for accurately characterizing the epidemic outbreak. We show that while a reduction to a single-layer model can accurately predict the epidemic characteristics when the network layers are purely Poisson, a departure from Poisson distribution can lead to incorrect predictions with single-layer models. Moreover, we show that the success of approaches that coalesce the multi-layer structure to an equivalent single-layer is critically dependent on the \emph{dispersion indices} of the network layers being perfectly matched. However, in practice,  different network layers (representing different congregate settings) are expected to have different structural characteristics, further highlighting the need for considering multi-layer network models for predicting the course of an outbreak. Our results further underscore the need for developing epidemiological models that account for heterogeneity among the pathogen variants as well as the contact network layers.

			\label{sec:Methods}
	\subsection*{Probability of Emergence}

	The first question that we investigate is whether a spreading process started by infecting a randomly chosen seed node with strain-1 causes an outbreak infecting a positive fraction of individuals, i.e., an outbreak of size $\Omega(n)$. Our results are based on multi-type branching processes \cite{ne:alexday, ne:pnas, haccou2005branching, mode1971multitype}. For computing the probability of emergence, we first define probability generating functions (PGFs) of the excess degree distribution: %\cite{newman2001random, newman2002spread}. We start by defining the following PGFs. 
% 	\begin{itemize}[noitemsep]
% 		\item 
		Let $g(z^f,z^w)$ denote the PGF for joint degree distribution of a randomly selected node (\emph{initial infective/ seed}) in the two network layers. This corresponds to the PGF for the probability distribution $p_{\boldsymbol{d}}=p_{d^f}^f    \cdot p_{d^w}^w $ and therefore,
		\begin{align}
			g(z^f,z^w)=\sum_{{\boldsymbol{d}}} p_{ \boldsymbol{d} } \left(z^f\right)^{d^f}\left(z^w\right)^{d^w}. \label{eq:small-g}
		\end{align}
% 		\item 
		For $a \in \{f,w\}$, we define,
		$G^a(z^f,z^w)$ as the PGF for excess joint degree distribution for the number of type-$f$ and type-$w$ contacts of a node reached by following a randomly selected type-$a$ edge (\emph{later-generation infective/ intermediate host}). While computing $G^a(z^f,z^w)$, we discount the type-$a$ edge that was used to infect the given node. We have
% 	\end{itemize}
	\begin{align}
		G^f(z^f,z^w)=\sum_{{\boldsymbol{d}}} \frac{d^f  p_{\boldsymbol{d}}}{\langle d^f \rangle } \left(z^f\right)^{d^f-1}\left(z^w\right)^{d^w} \label{eq:big-g-f},\\
		G^w(z^f,z^w)=\sum_{{\boldsymbol{d}}} \frac{d^w  p_{\boldsymbol{d}}}{\langle d^w \rangle } \left(z^f\right)^{d^f}\left(z^w\right)^{d^w-1}.\label{eq:big-g-w}
	\end{align}
	The factor  
	$d^f p_{\boldsymbol{d}}/ \langle d^f \rangle  $ (resp., $d^w p_{\boldsymbol{d}}/ \langle d^w \rangle  $) gives  the {\em normalized} probability that an edge of 
	type-$f$ (resp., type-$w$) is attached (at the other end) to a vertex
	with colored degree $\boldsymbol{d}=(d^f, d^w)$ \cite{newman2002spread}.
	
	Suppose, an arbitrary node $u$ carries strain-1 and transmits the infection to one of its susceptible neighbors, denoted as node $v$. Since there are two types of links/edges in the contact network and two types of strains circulating in the host population, there are four types of events that lead to the transmission of infection from node $u$ to $v$, namely, whether edge $(u,v)$ is 
	\begin{enumerate}[noitemsep]
		\item[(i)] type-$f$ and no mutation occurs in host $v$;
		\item[(ii)] type-$f$ and mutation to strain-2 occurs in host $v$;
		\item[(iii)]  type-$w$ and no mutation occurs in host $v$;
		\item[(iv)]  type-$w$ and mutation to strain-2 occurs in host $v$.
	\end{enumerate}
	In cases (i) and (iii) (resp., cases (ii) and (iv)) above, node $v$ acquires strain-1 (respectively, strain-2). For applying a branching process argument \cite{newman2001random, newman2002spread} and writing recursive equations using PGFs, it is crucial to keep track of both the types of edges used to transmit the infection and the types of strain acquired after mutation. Therefore, we keep a record of the number of newly infected individuals who acquire strain-1 or strain-2, and the type of edge through which they acquired the infection. We define the joint PGFs for transmitted infections over four random variables corresponding to the four infection events (i) - (iv) as follows.% with the subscript $1,2$ denoting the type of strain acquired by the host after mutation and the superscript $f,w$ denoting the type of edge through which the infection event occurred. 

% 	{small
%   \setlength{\abovedisplayskip}{6pt}
%   \setlength{\belowdisplayskip}{\abovedisplayskip}
%   \setlength{\abovedisplayshortskip}{0pt}
%   \setlength{\belowdisplayshortskip}{3pt}
%   \begin{align}
%  		&\gamma_1(z^f_1,z^f_2, z^w_1,z^w_2 ) \nonumber\\ 
% 		&= g \left(1-\tfone+\tfone \left( \sum_{j=1}^2 \mu^f_{1j} z^f_j\right), 1-\twone+\twone \left(\sum_{j=1}^2 \mu^w_{1j} z^w_j\right) \right),
%   \end{align}
% }%
% 	\begin{equation}
% \resizebox{.9\hsize}{!}{ $g \left(1-\tfone+\tfone \left( \sum_{j=1}^2 \mu^f_{1j} z^f_j\right), 1-\twone+\twone \left(\sum_{j=1}^2 \mu^w_{1j} z^w_j\right) \right)$}
% \end{equation}
	\begin{align}
		&\gamma_1(z^f_1,z^f_2, z^w_1,z^w_2 )= \nonumber\\ 
		& g \left(1-\tfone+\tfone \left( \sum_{j=1}^2 \mu^f_{1j} z^f_j\right), 1-\twone+\twone \left(\sum_{j=1}^2 \mu^w_{1j} z^w_j\right) \right). \nonumber
	\end{align}
	For $a \in \{f,w\}$ and $i \in \{1,2 \}$, denote
	\begin{align}
		&\Gamma^a_i(z^f_1,z^f_2, z^w_1,z^w_2 )=\nonumber\\ 
		& G^a \left(1-T^f_i+T^f_i \left(\sum_{j=1}^2 \mu^f_{ij} z^f_j \right), 1-T^w_i+T^w_i \left(\sum_{j=1}^2 \mu^w_{ij} z^w_j \right)\right). \nonumber 
	\end{align}
	We %derive the functions $\gamma_1$ and $\Gamma^a_i$ and 
	show that the quantity $\gamma_1(z^f_1,z^f_2, z^w_1,z^w_2 )$ represents the PGF for the number of infection events of each type induced among the neighbors of a seed node when the seed node is infected with strain-1; see \emph{{\color{blue}SI Appendix 1.A}}. Furthermore, for $a \in \{f,w\}$ and $i \in \{1,2\}$,  we show that $\Gamma^a_i(z^f_1,z^f_2, z^w_1,z^w_2)$ is the PGF for number of infection events of each type caused by a {\em later-generation} infective (i.e., a typical intermediate host in the process) that received the infection through a type-$a$ edge and carries strain-$i$. Building upon the PGFs for the infection events caused by the seed and later-generation infectives, our first main result characterizes the probability of emergence when the outbreak starts at an arbitrary node infected with strain-1.\\ %{\color{red}Figure for \emph{\color{blue}SI Appendix}}
	%is the PGF for the Excess \emph{occupied/infected} degree of a node reached by following a randomly selected edge PGF for the number of infections of each type transmitted by a {\em later-generation} infective of type-$i$
%	[NOTE- the PNAS template is not supporting the typical theorem environments]
	\noindent\emph{Theorem 1{ (Probability of Emergence):} It holds that
		\begin{align}
			%Probability of extinction  
			\pr {[\rm Emergence]} =1- \gammaoneQ,			\label{ne:th:prob}
		\end{align}
		where, $\left(q^f_1,q^f_2, q^w_1,q^w_2\right)$ are the smallest non-negative roots of the fixed point equations:
		\begin{align}
			\qfone&=\GammafoneQ\label{eq:ne:rec1}\\
			\qftwo&=\GammaftwoQ\label{eq:ne:rec2}\\
			\qwone&=\GammawoneQ\label{eq:ne:rec3}\\
			\qwtwo&=\GammawtwoQ\label{eq:ne:rec4}.
		\end{align}}Here, for $a \in \{w,f\}$ and $i \in \{1,2\}$, the term $q^a_i$ can be interpreted as the probability of extinction starting from one later-generation infective carrying strain-$i$ (after mutation) which was infected through a type-$a$ edge; see \emph{{\color{blue}SI Appendix 1.A}} for a detailed proof. Therefore, the probability of emergence of an epidemic is given by the probability that at least one of the infected neighbors of the seed triggers an unbounded chain of transmission events. We note that Theorem~1 provides a strict generalization for the probability of emergence of multi-strain spreading on a single layer \cite{ne:alexday} and we can recover the probability of emergence for the case of single layer by substituting $\pmb{T}^f=\pmb{T}^w$ and $\pmb{\mu}^f=\pmb{\mu}^w$ in Equations~\eqref{eq:ne:rec1}-~\eqref{eq:ne:rec4}.
		
		%The theorem states that for $a \in \{w,f\}$ and $i \in \{1,2\}$, , denoted $q^a_i$, is the smallest non-negative root of the equation $q^a_i = \Gamma^a_i \left(q^f_1,q^f_2, q^w_1,q^w_2\right)$ solved simultaneously for $i=1,2$. 

	%We provide in 
%		\label{ne:th:prob}
% 	\end{thm}
	% Finally, the overall extinction probability is given by $g \left(1-T_i+T_i\sum_{j=1}^m \mu_{ij} q_j \right)$ if the whole process starts with an initial infective of type-$i$."
	
	%\textbf{Proof of Theorem~\ref{ne:th:prob} }
	\subsection*{Epidemic Threshold}
	Next, we characterize the {epidemic threshold}, which defines a boundary of the region in the parameter space inside which the outbreak always dies out after infecting only a finite number of individuals; while, outside which, there is a positive probability of a positive fraction of infections. The epidemic threshold is commonly studied  as a metric to characterize and epidemic and ascertain risk factors \cite{rho-survey}.
	%Let $k_f$ and $k_w$ be random variables
	%independently drawn from 
	Let $\lambda_f$ and $ \lambda_w$ denote the first moments of
	the distributions $\{p_{d^f}^f   \}  $ and
	$\{p_{d^w}^w\} $, respectively. Let  $\langle d_f^2 \rangle $ and  $\langle d_w^2 \rangle $ denote the corresponding second moments for distributions $\{p_{d^f}^f\}$ and
	$\{p_{d^w}^w\}$. Further, define $\beta_f$ and $\beta_w$ as the mean of the excess degree distributions respectively in the two layers. We have
	\begin{equation}
		\label{eq:betas_new_def}
		\beta_f:= \frac{ \langle d_f^2 \rangle  -\lambda_f }{\lambda_f} \quad \textrm{and}
		\quad \beta_w:= \frac{ \langle d_w^2 \rangle  -\lambda_w }{\lambda_w}.
	\end{equation}
% 		\label{ne:th:thresh}
		\noindent\emph{Theorem 2 (Epidemic Threshold): For}
		\begin{align}
			\boldsymbol{J}= \left[   \begin{matrix}
				\tfone \mufoneone \beta_f &  \tfone \mufonetwo\beta_f &\twone \muwoneone \lambda_w & \twone \muwonetwo \lambda_w\\ 
				\tftwo \muftwoone \beta_f &  \tftwo \muftwotwo\beta_f & \twtwo \muwtwoone \lambda_w & \twtwo \muwtwotwo  \lambda_w\\
				\tfone \mufoneone\lambda_f & \tfone \mufonetwo \lambda_f & \twone \muwoneone \beta_w & \twone \muwonetwo \beta_w\\ 
				\tftwo \muftwoone \lambda_f &   \tftwo \muftwotwo \lambda_f &\twtwo \muwtwoone\beta_w &\twtwo \muwtwotwo \beta_w 
			\end{matrix} \right],
			\label{eq:J_p}
		\end{align}
		\emph{let
		$\sigma(\boldsymbol{J})$ denote the spectral radius of $\sigma(\boldsymbol{J})$. The epidemic threshold is given by $\sigma(\boldsymbol{J})=1$.}\\ 
		The above theorem states that the epidemic threshold is tied to the spectral radius of the Jacobian matrix $J$, i.e., if $\sigma(\boldsymbol{J}) > 1$ then an epidemic occurs with a positive probability, whereas if $\sigma(\boldsymbol{J}) \leq 1$ then with high probability the infection causes a \emph{self-limited} outbreak, where the fraction of infected nodes vanishes to 0 as $n \rightarrow \infty$.% where the number of infected individuals is $o(n)$.%infecting a vanishingly small number of individuals $o(n)$.
%	\end{thm}
	 The matrix $\boldsymbol{J}$ is obtained while determining the stability of the fixed point of the recursive equations in Theorem~1 by linearization around $\qfone=\qftwo=\qwone=\qwtwo=1$ (\emph{\color{blue}SI Appendix 1.B}).% yields the Jacobian matrix in Theorem~2.
	 
 We note that when the mutation matrix is \emph{indecomposable}, meaning that each type of strain eventually may have lead to the emergence of any other type of strain with a positive probability, the threshold theorem for multi-type branching processes \cite{ne:alexday} guarantees if $\sigma(\boldsymbol{J})\leq 1$, then $q^a_i=1$; whereas if $\sigma(\boldsymbol{J})> 1$, then $0 \leq q^a_i <1$, where $ i \in \{1,2\}$ and $a \in \{f,w\}$. For \emph{decomposable} processes, the threshold theorem \cite{ne:alexday} guarantees extinction ($q^a_i=1$) if $\sigma(\boldsymbol{J})\leq 1$; however the uniqueness of the fixed-point solution does not necessarily hold when $\sigma(\boldsymbol{J})>1$.
%{\color{blue} Note on decomposable/indecomposable mutation matrices based on \cite{ne:alexday} \emph{\color{blue}SI Appendix}.}

 Our next result provides a decoupling of the epidemic threshold into causal factors pertaining pathogen and mutation, and structural properties of different layers in the contact network.  \\
%{\color{red} (I think the following sentence needs to be pushed to a new line as it is not related)}

\noindent\emph{Lemma 1: When ${\twone}/{\tfone} ={\twtwo}/{\tftwo}=c$, where $c>0$, and let $\pmb{\mu}=\pmb{\mu}^f=\pmb{\mu}^w$,
we get,
	\begin{align}
		\sigma(\pmb{J}) = \sigma \left(	\left[   \begin{matrix}
			\beta_f &  c \lambda_w\\\ 
			\lambda_f &  c \beta_w
		\end{matrix} \right] \right) \times \sigma(\pmb{T^f} \pmb{\mu}). \label{eq:spec-prod}
	\end{align}}
Lemma 1 follows from the observation that with ${\twone}/{\tfone} ={\twtwo}/{\tftwo}=c$, we can express $\pmb{J}$ as a Kronecker product of two matrices (denoted by $\otimes $), as below.	\begin{align}
		\boldsymbol{J}&= \left[   \begin{matrix}
			\tfone \muoneone \beta_f &  \tfone \muonetwo\beta_f &\twone \muoneone \lambda_w & \twone \muonetwo \lambda_w\\ 
			\tftwo \mutwoone \beta_f &  \tftwo \mutwotwo\beta_f & \twtwo \mutwoone \lambda_w & \twtwo \mutwotwo  \lambda_w\\
			\tfone \muoneone\lambda_f & \tfone \muonetwo \lambda_f & \twone \muoneone \beta_w & \twone \muonetwo \beta_w\\ 
			\tftwo \mutwoone \lambda_f &   \tftwo \mutwotwo \lambda_f &\twtwo \mutwoone\beta_w &\twtwo \mutwotwo \beta_w 
		\end{matrix} \right]  \nonumber\\
% 		& =\left[   \begin{matrix}
% 			\tfone \muoneone \beta_f &  \tfone \muonetwo\beta_f &c\tfone \muoneone \lambda_w & c\tfone \muonetwo \lambda_w\\ 
% 			\tftwo \mutwoone \beta_f &  \tftwo \mutwotwo\beta_f & c\tftwo \mutwoone \lambda_w & c\tftwo \mutwotwo  \lambda_w\\
% 			\tfone \muoneone\lambda_f & \tfone \muonetwo \lambda_f & c\tfone \muoneone \beta_w & c\tfone \muonetwo \beta_w\\ 
% 			\tftwo \mutwoone \lambda_f &   \tftwo \mutwotwo \lambda_f &c\tftwo \mutwoone\beta_w &c\tftwo \mutwotwo \beta_w 
% 		\end{matrix} \right]  \nonumber\\
		&= 	\left[   \begin{matrix}
			\beta_f &  c \lambda_w\\\ 
			\lambda_f &  c \beta_w
		\end{matrix} \right] \otimes (\pmb{T^f} \pmb{\mu}).
	\end{align}
We note that the first assumption ${\twone}/{\twtwo} ={\tfone}/{\tftwo}$ is consistent with scenarios where the ratio of the transmissibility of the two strains in each layer is expected to be a property of the pathogen and not the contact networks. This assumption is supported by the typical modeling assumption \cite{aleta2020modelling} that social distancing measures such as increasing distance between individuals lead to a reduction in the transmissibility of the disease by a specific coefficient for the entire network layer. And therefore, when each network layer has specific restrictions in place (and corresponding coefficients for reduction in transmissibility), the ratio of the transmissibility of the two strains in each layer ends up being a property of the heterogeneity in the strains. The second assumption ($\pmb{\mu}^f=\pmb{\mu}^w$) in Lemma 1 is motivated by the assumption that mutations occur \emph{within} individual hosts, which is typical to multi-strain spreading models; see \cite{ne:alexday} and the references therein.

% \textbf{A unified anaylsis for spectral radius}: 
We note that the decoupling obtained through \eqref{eq:spec-prod} reveals the delicate interplay of the network structure and the transmission parameters in determining the threshold for emergence of an epidemic outbreak. Lastly, we observe that Lemma 1 provides a unified analysis for the spectral radius including the case with a single-strain or a single-layer. 
 %The probability equations  can be derived by substituting $q^f_1=q^w_1$ and $q^f_1=q^w_1$.
For the multi-strain spreading on a single-layer network, the spectral radius can be derived by substituting $T^f_i=T^w_i$ in \eqref{eq:spec-prod} and setting mean degree of one of the layers as 0, for instance, setting $\lambda_w=\beta_w=0$, yielding the epidemic threshold, denoted as $\rho^{\rm MS-SL}$,
\begin{align}
    \rho^{\rm MS-SL}=\beta_f \times \sigma(\pmb{T^f} \pmb{\mu}),\label{eq:rho-MS-SL}
\end{align}
where $\beta$ corresponds to the mean of the excess degree distribution for the single-layered contact network. For the case of the spread of a single strain on a multi-layer contact network, we substitute $\tfone=\tftwo$ in \eqref{eq:spec-prod}, which implies $\rho(\pmb{T^f}\pmb{\mu})=T^f \rho (\pmb{\mu})=T^f$, yielding the epidemic threshold, denoted as $\rho^{\rm SS-ML}$,
	\begin{align}
	 \rho^{\rm SS-ML} &= \sigma \left(	\left[   \begin{matrix}
			\beta_f &  c \lambda_w\\\ 
			\lambda_f &  c \beta_w
		\end{matrix} \right] \right) \times T^f.\label{eq:rho-SS-ML}%\\
% 		&=\rho \left(	\left[   \begin{matrix}
% 			T^f \beta_f &  T^w \lambda_w\\\ 
% 			T^f \lambda_f &  T^w \beta_w
% 		\end{matrix} \right] \right).
	\end{align}
It is easy to verify that the spectral radius as obtained from \eqref{eq:rho-MS-SL} and \eqref{eq:rho-SS-ML} is consistent with the results in \cite{ne:pnas} and \cite{ne:alexday}.
% single-strain model
% , thus enabling better identification and control of different control factors.
% interplay
% gives a more interpretable version of the epidemic threshold separately accounting for contributions from the network and transmission parameters.
% It enables us to identify and control the 
% \textbf{Decoupling into network and transmission parameters}\\
%  The decoupling of a spectral radius into a product of spectral radii of two matrices- one dependent on the layer parameters and the second on the transmission and mutation parameters - highlight as a key result
% delicate interplay of 
\subsection*{Mean Epidemic Size}
	\label{sec:ne:size}
	Next, we compute the mean epidemic size and the mean fraction of nodes infected by each type of strain. The knowledge of the fraction of individuals infected by each strain is vital for cases when different pathogen strains have different transmissibility and virulence. In such cases, predicting the expected fraction of the population hit by the more severe strain can help scale healthcare resources in time.

	For computing the mean epidemic size, we consider the zero-temperature random-field Ising model on Bethe lattices \cite{ne:Sethna1993}, as done in \cite{ne:pnas}. We refer to a node as being {\em active} if it is infected with either of the two strains (strain-1 or strain 2), and {\em inactive} otherwise. Since $\mathbb{H}$ is locally tree-like \cite{Soderberg3}, we consider the following hierarchical structure, such that at the top level, there is a single node (the {\em root}). The probability that an arbitrarily chosen root node is infected with strain 1 (resp., strain 2), gives mean value for fraction of individuals infected by strain 1 (resp., strain 2). Let $Q_1$ (resp., $Q_2$) denote the probability that the root node is active and carries strain-1 (resp., strain-2). We label the levels of the tree from level $\ell=0$ at the bottom to level $\ell = \infty$ at the top, i.e., the root.
	We assume that {\em co-infection} is not possible, hence a node that receives $x^f_1$ (resp., $x^w_1$) infections of strain-$1$ through type-$f$ (resp., type-$w$) links, and  $x^f_2$ (resp., $x^w_2$) infections of strain-$2$ through type-$f$ (resp., type-$w$) links, then it becomes infected by strain-$i$  with probability %{\color{red} (I recommend writing the following equation in its own line (display style))} 
 $$\frac{\mu^f_{1i} x^f_1+\mu^f_{2i}x^f_2+\mu^w_{1i}x^w_1+\mu^w_{2i}x^w_2}{x^f_1+x^f_2+x^w_1+x^w_2},$$ where $i \in \{1,2\}$. 
	For $a \in \{f,w\}$, and $i \in \{1,2\}$, let $q^a_{\ell+1,i}$ denote the probability that a node at level $\ell+1$ is active,  carries strain-$i$ and is connected to a node at level $\ell+2$ through a type-$a$ edge. Our next result characterizes the mean fraction of individuals infected by each type of strain during an epidemic outbreak. As a crucial step towards deriving the mean fraction of infected individuals, we first show that for $i=1,2$, we have
	\begin{align}
		q^f_{\ell+1, i} &= \sum_{\boldsymbol{d} } \frac{d^f  p_{\boldsymbol{d}}}{\langle d^f \rangle }  f_i(q^f_{\ell,1},q^f_{\ell,2},q^w_{\ell,1},q^w_{\ell,2},d^f-1, d^w),	\label{ne:eq:size-rec-levels-f}\\
		q^w_{\ell+1, i} &=  \sum_{\boldsymbol{d} } \frac{d^w  p_{\boldsymbol{d}}}{\langle d^w \rangle }  f_i(q^f_{\ell,1},q^f_{\ell,2},q^w_{\ell,1},q^w_{\ell,2},d^f, d^w-1)
		\label{ne:eq:size-rec-levels-w},
	\end{align}
%We present a proof of \eqref{ne:eq:size-rec-levels-w} and \eqref{ne:eq:size-rec-levels-f}  in the \emph{\color{blue}SI Appendix}. 
For $a\in \{f,w\}, i \in \{1,2\}$, let $q^a_{\infty,i}$ denote the limit of $q^a_{\ell,i}$ as $\ell \rightarrow \infty$.\\	
	\emph{Theorem 3 (Epidemic Size)
	For $i=1,2$, we have
		\begin{align}
		Q_i &= \sum_{{\boldsymbol{d}}}p_{\boldsymbol{d}} f_i(q^f_{\infty,1},q^f_{\infty,2},q^w_{\infty,1},q^w_{\infty,2},d^f, d^w),	\label{ne:eq:size-rec-root}
		% Q_2 &= \sum_{p_{\boldsymbol{d}}}p_{\boldsymbol{d}} f(q^f_{\infty,1},q^f_{\infty,2},q^w_{\infty,1},q^w_{\infty,2},d^f, d^w),
	\end{align}
	with the mean epidemic size $$Q = Q_1+Q_2,$$ where $f_i(u^f_1,u^f_2,u^w_1,u^w_2, z^f, z^w)$ is given in {\color{blue}SI {Appendix} 1.C}}.\\% \eqref{ne:eq:size-rec-f}.}\\
	 Here, $f_i(u^f_1,u^f_2,u^w_1,u^w_2, z^f, z^w)$ denotes the probability that an arbitrary node $u$ at level $\ell+1$ gets infected with strain-$1$ through neighbors in level $\ell$ such that there are $z^f$ and $z^w$ neighbors of node $u$ in layers $f$ and $w$ respectively. A precise definition of 
 $f_i(u^f_1,u^f_2,u^w_1,u^w_2, z^f, z^w)$ and a proof of Theorem 3 is presented in the \emph{\color{blue}SI Appendix 1.C}. As for the previous Theorems, we observe that Theorem 3 collapses to the multi-strain spreading on a single network layer by substituting the transmissibilities of the two network layers as being equal.

 	\subsection*{Experimental Evaluation}
	\label{sec:exp-simulations}
	In this section, we present numerical studies on different contact structures and transmission patterns. For our simulations, we focus on the setting where the fitness landscape consists of two types of strains and two types of network layers. The two network layers are independently generated using the configuration model after sampling degree sequences from the distributions for the two layers $\{p^f_k, \: k=0,1,\ldots\}$ and $\{p_k^w, \: k=0,1,\ldots\}$. We first present results for the case when the network layers follow a Poisson degree distribution. Next, we consider the power law degree distribution  which is widely used \cite{powerlaw-empirics} in modeling the structure of several real-world networks including social networks. Further, to account for mitigation measures that limit the number of people who can congregate in the different layers, we let the degree distributions for the two layers follow the power law degree distribution with exponential cutoff for both layers; (\emph{{\color{blue}SI Appendix 1.D}}). %{\color{red} (curious what ``gathering limits" means)}
	
	The spreading process is initiated when a randomly chosen node is selected as the seed carrying strain-1  (Figure~\ref{fig:ne:txmodel}). In subsequent time-steps, each node independently infects their neighbors with a transmission probability that depends on both the type of strain carried and the nature of link through which contact occurs. After infection, the pathogen mutates within the hosts with probabilities given by the mutation matrices. In cases where a susceptible node comes in contact with multiple infectious neighbors, we resolve exposure to multiple infections by assigning the probability of acquiring each strain as the fraction of exposures received for that strain. In particular, if a node receives $x^f_1$ (resp., $x^w_1$) infections of strain-$1$ through type-$f$ (resp., type-$w$) links, and  $x^f_2$ (resp., $x^w_2$) infections of strain-$2$ through type-$f$ (resp., type-$w$) links,  with probability $\frac{\mu^f_{1i} x^f_1+\mu^f_{2i}x^f_2+\mu^w_{1i}x^w_1+\mu^w_{2i}x^w_2}{x^f_1+x^f_2+x^w_1+x^w_2}$, for $i \in \{1,2\}$ it acquires strain-$i$, which it spreads to its neighbors. The process reaches a steady state and terminates when no new infections are possible.  Throughout, we let $Q$ denote the mean epidemic size and $Q_1$ and $Q_2$, respectively denote the final fraction of individuals infected by each strain in the steady state.
	
\begin{figure}[t]
		\centering
		\includegraphics[scale=0.27]{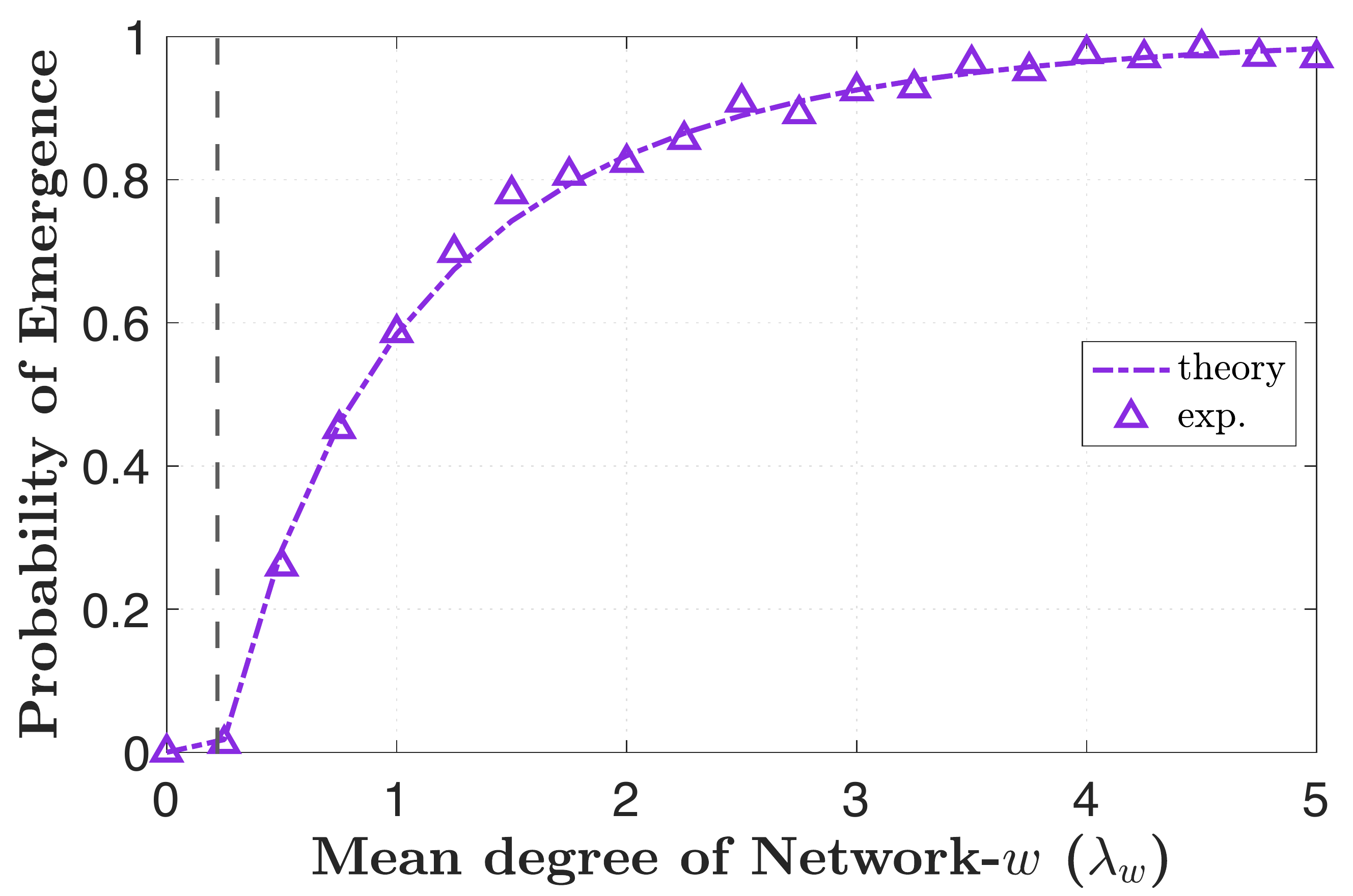}
				\includegraphics[scale=0.27]{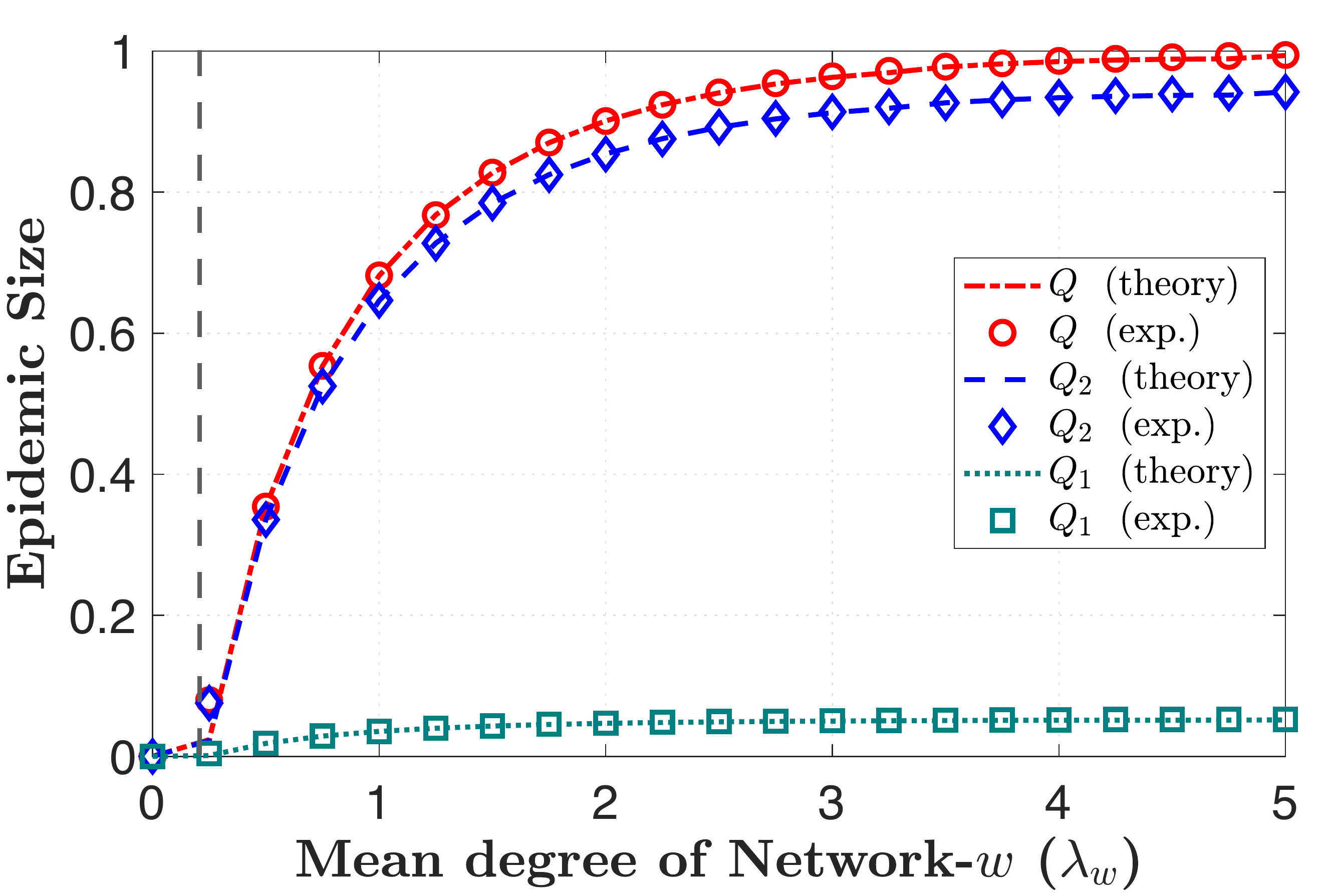}
		\caption{\sl Probability of emergence and the fraction of individuals infected with each strain $(Q_1, Q_2)$ for a contact network comprising two Poisson layers with mean degrees $\lambda_f$ and $\lambda_w$ respectively. We set $\tfone =0.6, \tftwo = 0.8 $, $\twone=0.7, \twtwo=0.9, \mufoneone=\muwoneone=0.1$, $\muftwotwo=\muwtwotwo=0.95$, and $n=10000$. We set $\lambda_f=1$ and vary $\lambda_w$ in $[0,5]$ . The theoretical probability, epidemic size and transition points are derived from Theorems~1-3, respectively, while the experimental results correspond to averaging over 500 independent experiments. } 
		\label{fig:ne:plot1}
	\end{figure}

	Next, we compare our analytical results for the probability of emergence and expected epidemic size (Theorems~1-3) with empirical values obtained by simulating the spread of infection over multiple independent experiments. We consider a contact network where the degree distribution for each layer is Poisson with parameters  $\lambda_f$ and $\lambda_w$, respectively. To model scenarios where there is a risk of the emergence of a new, more transmissible strain (strain-2) starting from strain-1, we set $\tfone =0.6, \tftwo = 0.8 $, $\twone=0.7, \twtwo=0.9, \mufoneone=\muwoneone=0.1$, and $\muftwotwo=\muwtwotwo=0.95$ and we fix the number of nodes $n=10000$. We plot the probability of emergence and epidemic size averaged over 500 independent experiments in Figure~\ref{fig:ne:plot1}. We indicate the epidemic threshold as the vertical dashed line where we observe a phase transition, with the probability and expected epidemic size sharply increasing from zero to one as the epidemic threshold is exceeded. We plot the expected fraction of individuals infected by each strain ($Q_1$ and $Q_2$). The total epidemic size $Q$ is the sum of the fraction of individuals infected by each strain ($Q=Q_1+Q_2$). We observe a good agreement between the experimental results and the theoretical predications given through Theorems 1-3. % {\color{red} (I think a conclusion is needed here. Something like we observe a good match, etc.)}.  
			\begin{figure}[t]
		\centering
		\includegraphics[scale=0.33]{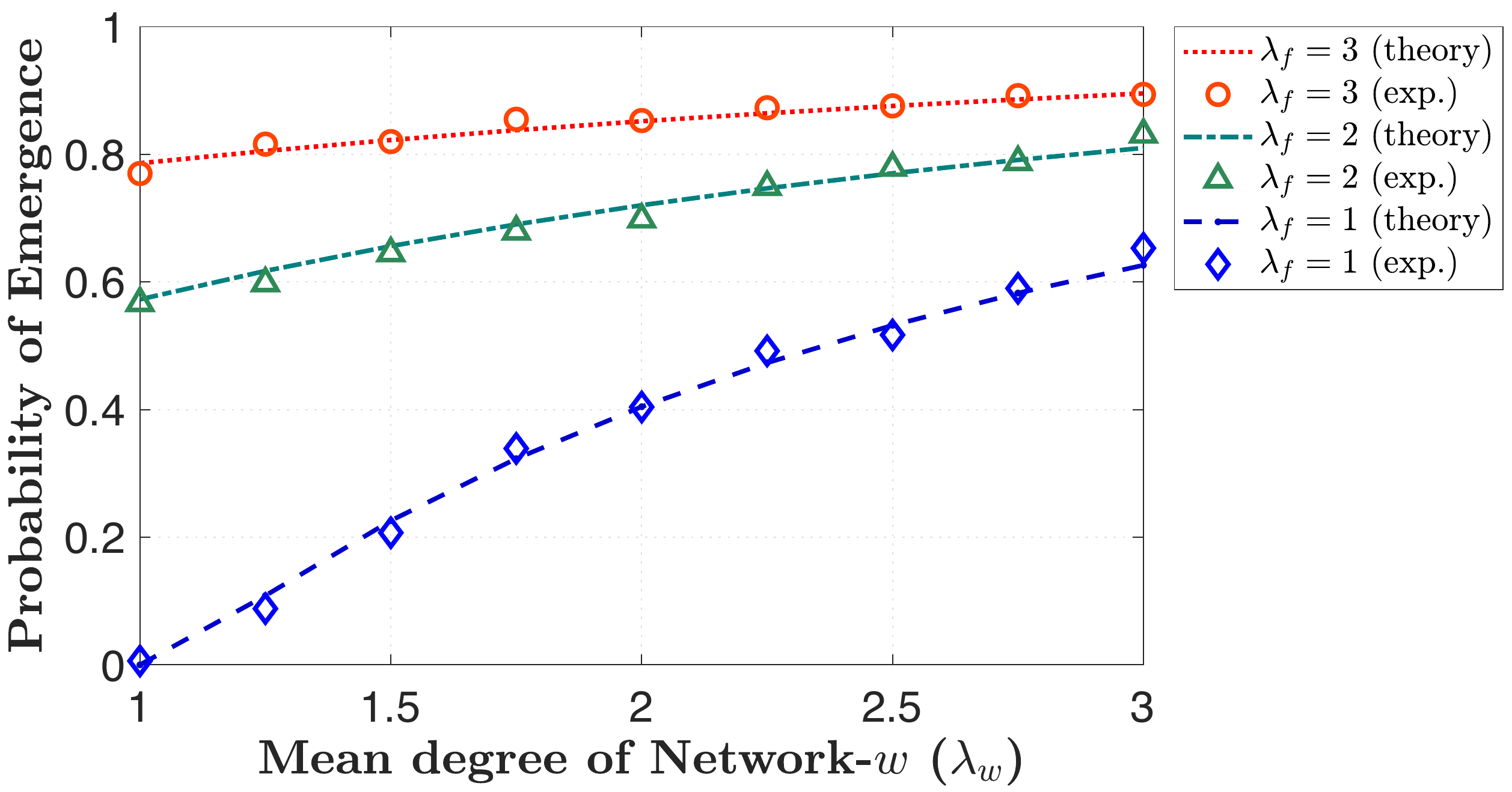}
				\includegraphics[scale=0.33]{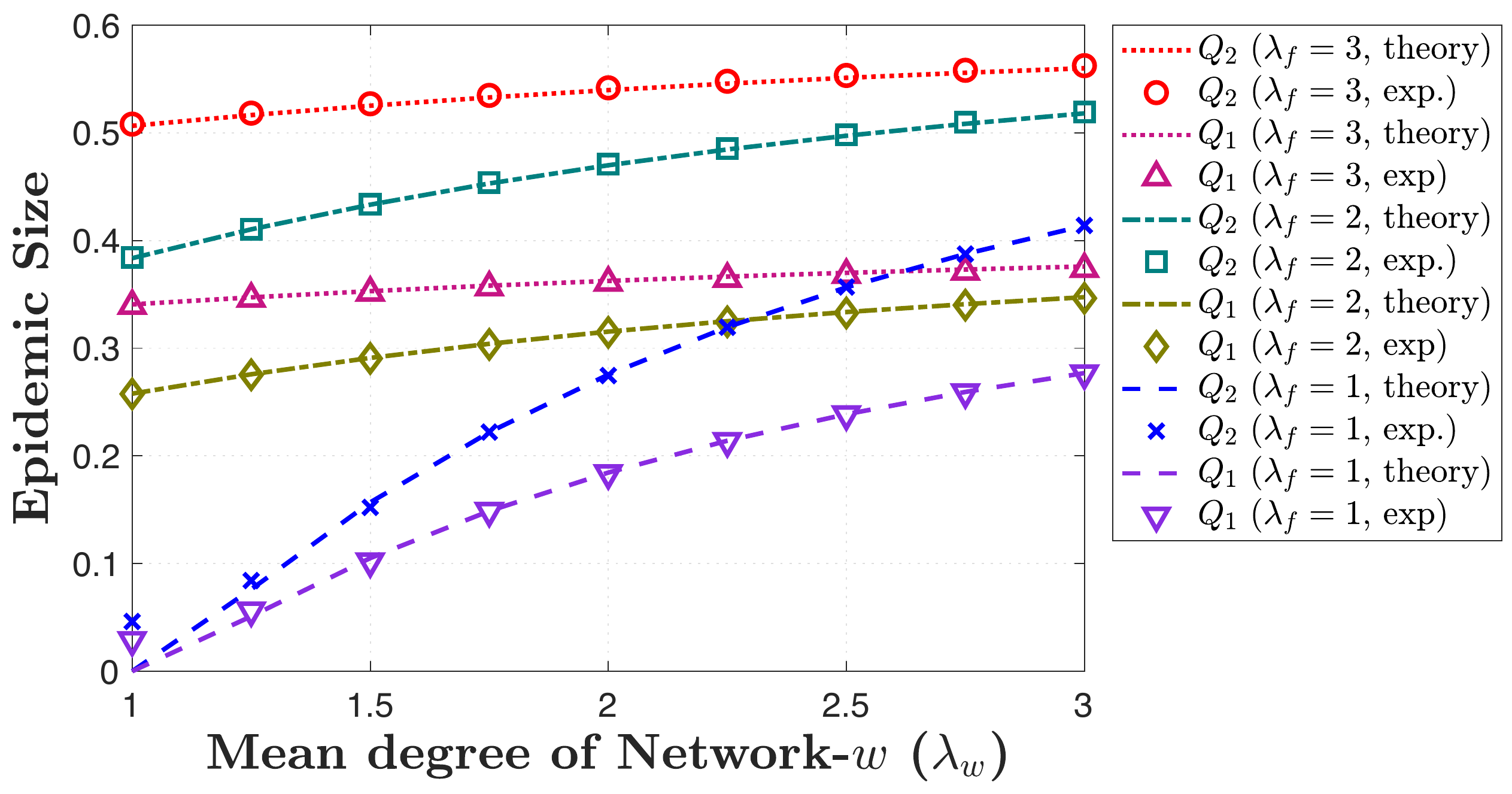}
		\caption{\sl Probability of emergence and the expected epidemic size for a contact network comprising two Poisson layers with mean degrees $\lambda_f$ and $\lambda_w$ respectively. The fixed parameters are $\tfone =0.5, \tftwo = 0.7 $, $\twone=0.3, \twtwo=0.4, \mufoneone=\muwoneone=0.2$, $\muftwotwo=\muwtwotwo=0.5$, $n=10000$, and we vary $\lambda_f$ and $\lambda_w$ in the interval $[1,3]$.} 
		\label{fig:ne:plot3}
	\end{figure} 

To demonstrate the impact of increasing edge density of the contact network, we vary the mean node degree while keeping transmission and mutation parameters fixed. In Figure~\ref{fig:ne:plot3}, we consider the case where the degree distribution for network layers are Poisson and we vary the mean degrees of the two layers. We consider $n=10000$, and consider a different initialization for the transmission parameters with $\tfone =0.5, \tftwo = 0.7 $, $\twone=0.3, \twtwo=0.4, \mufoneone=\muwoneone=0.2$, and  $\muftwotwo=\muwtwotwo=0.5$. To see the impact of the edge density of the two layers on the epidemic characteristics, we vary $\lambda_f$ and $\lambda_w$ in $[1,3]$. The probability of emergence and epidemic size are averaged over 1000 independent experiments. We observe a good agreement between our analytical results in Theorems 1-3 and simulations in Figures~\ref{fig:ne:plot1} and \ref{fig:ne:plot3}.
%{\color{blue}\emph{\color{blue}SI Appendix}- power law}
% 	{\color{red} 1 figure}
% 	{\em figure caption}

\section*{Discussion}
	\subsection*{Joint impact of layer openings and mutations}
			Next, we discuss the interplay of layer openings and mutations on the probability of emergence of epidemic outbreaks. We consider the case where the fitness landscape consists of two strains. The process starts when the population is introduced to the first strain (strain-1) which is moderately transmissible and initially dominant in the population. In contrast, the other strain (strain-2) is highly transmissible and initially absent in the population but has the risk of emerging through mutations in strain-1. For modeling mutations that occur within hosts, we assume the mutation probabilities depend on the type of strain but not on the type of link over which the infection was transmitted. We let $\pmb{\mu}$ denote the mutation matrices in the two layers ($\pmb{\mu}=\pmb{\mu}^f=\pmb{\mu}^w$), and $\mutwotwo \rightarrow 1$, wherein with high probability, once the pathogen mutates to strain-2, it does not mutate back to strain-1. 
In particular, we consider the one-step irreversible  mutation and transmission matrices as below:
\begin{align}
		\pmb{\mu}= \left[   \begin{matrix}
			\mu_{11} &  1-\mu_{11}\\ 
			0 &  1
		\end{matrix} \right], 0<\muoneone<1 \label{eq:onesteprev}
	\end{align}
		\begin{align}
	\left[   \begin{matrix}
		\twone &  0\\ 
		0 &  \twtwo
	\end{matrix} \right]= c\left[   \begin{matrix}
		\tfone & 0\\ 
		0 &  \tftwo
	\end{matrix} \right], c\geq1;~~ T^w_1<T^w_2 {\rm{~and~}} \tfone<T^f_2 \label{eq:transmission_coupling}.
\end{align}
			The above mutation and transmission parameters in \eqref{eq:onesteprev} correspond to the \emph{one-step irreversible} mutation scheme, which is used widely \cite{antia2003role, ne:alexday} to model scenarios where a simple change is required for the contagion to evolve to a highly transmissible variant. 
 We first derive a result characterizing the epidemic threshold for one-step irreversible mutations.\\
\emph{Lemma 2: For the one-step irreversible mutation matrix given by \eqref{eq:onesteprev} and transmission parameters satisfying \eqref{eq:transmission_coupling}, the epidemic threshold does not depend on $\mu_{11}$ or $\tfone$. Specifically,	\begin{align}
		\sigma(\pmb{J}) = \tftwo \times \sigma \left(	\left[   \begin{matrix}
			\beta_f &  c \lambda_w\\\ 
			\lambda_f &  c \beta_w
		\end{matrix} \right] \right) . \label{eq:spec-prod-irrev}
	\end{align}
}
A proof for Lemma~2 is provided in \emph{\color{blue}SI Appendix 2.A}.

In cases where the initial strain by itself is not transmissible enough to cause an epidemic and the course of the epidemic is tied to the emergence of a highly transmissible strain, we find it useful to derive the following epidemiological parameter. In such scenarios, in addition to the probability of emergence, it is useful to derive the probability that at least one mutation to strain-2 occurs in the chain of infections initiated by the introduction of strain-1 to a seed node. Note that the above quantity is different from the probabilities given by the mutation matrix, which defines the probability of mutation to a different strain (within each host) after every transmission event. \\
\emph{Lemma 3: For the one-step irreversible mutation matrix given by \eqref{eq:onesteprev} and transmission parameters satisfying \eqref{eq:transmission_coupling}, we have
\begin{align}
  	&\pr[ \rm{at~least~1~mutation~to~strain-2}~]\nonumber\\ &=1-g(1-T^f_1+T^f_1\mu_{11}q^f,1-T^w_1+T^w_1\mu_{11}q^w),  \label{ne:eq:bound-rec-0}
\end{align}
where,
	\begin{align}
		q^f &= G^f(1-T^f_1+T^f_1\mu_{11}q^f,1-T^w_1+T^w_1\mu_{11}q^w)\label{ne:eq:bound-rec-1},\\
		q^w &= G^w(1-T^f_1+T^f_1\mu_{11}q^f,1-T^w_1+T^w_1\mu_{11}q^w)\label{ne:eq:bound-rec-2}.
	\end{align}
}
Here, for $a\in \{f,w\}$, $q^a$ corresponds to the probability of there being no mutation to strain-2 in the chain of infections emanating from a later generation infective that was infected through a type-$a$ edge; details in \emph{\color{blue}SI Appendix 2.B}.

	 	 			\begin{figure}[t]
		\centering
		\includegraphics[scale=0.15]{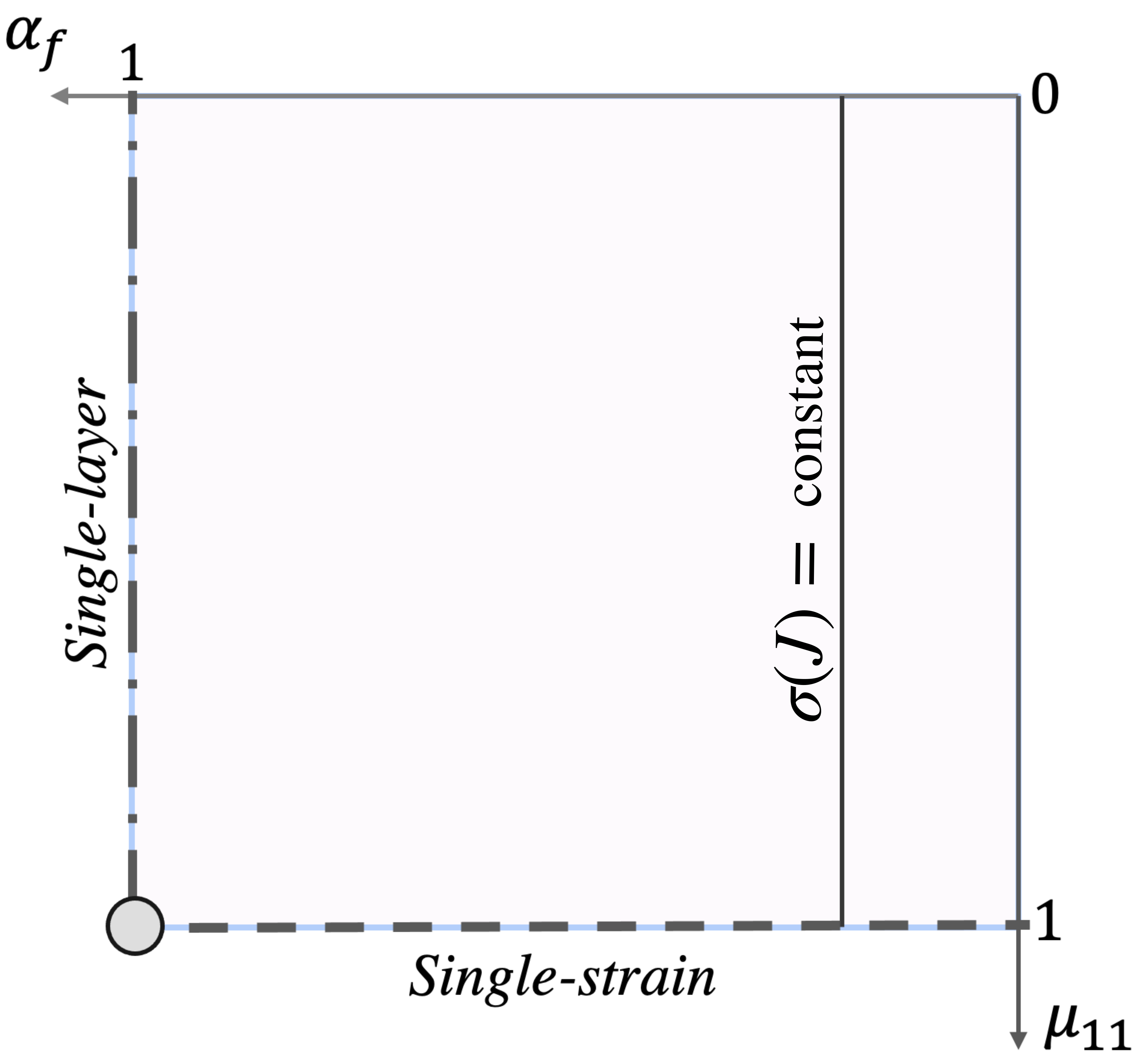}
		\caption{\sl A parameterization of the multi-layer multi-strain model as it transitions from the presence of one layer to two layers and one strain to two strains. Here, $\alpha_f \rightarrow 1$ corresponds to the case when there are no edges in layer-$f$ and the contact network $\mathbb{H}$ effectively has a single-layer $\mathbb{W}$, whereas $\muoneone \rightarrow 1$ corresponds to the case where no mutations to a different strain occur starting from strain-1 with high probability. The vertical lines correspond to contours with a constant epidemic threshold $\sigma(\pmb{J})$ (Lemma 2). } 
		\label{fig:ne:surf-domain}
	\end{figure} 
		\begin{figure*}[t]
		\centering
			\includegraphics[scale=0.30]{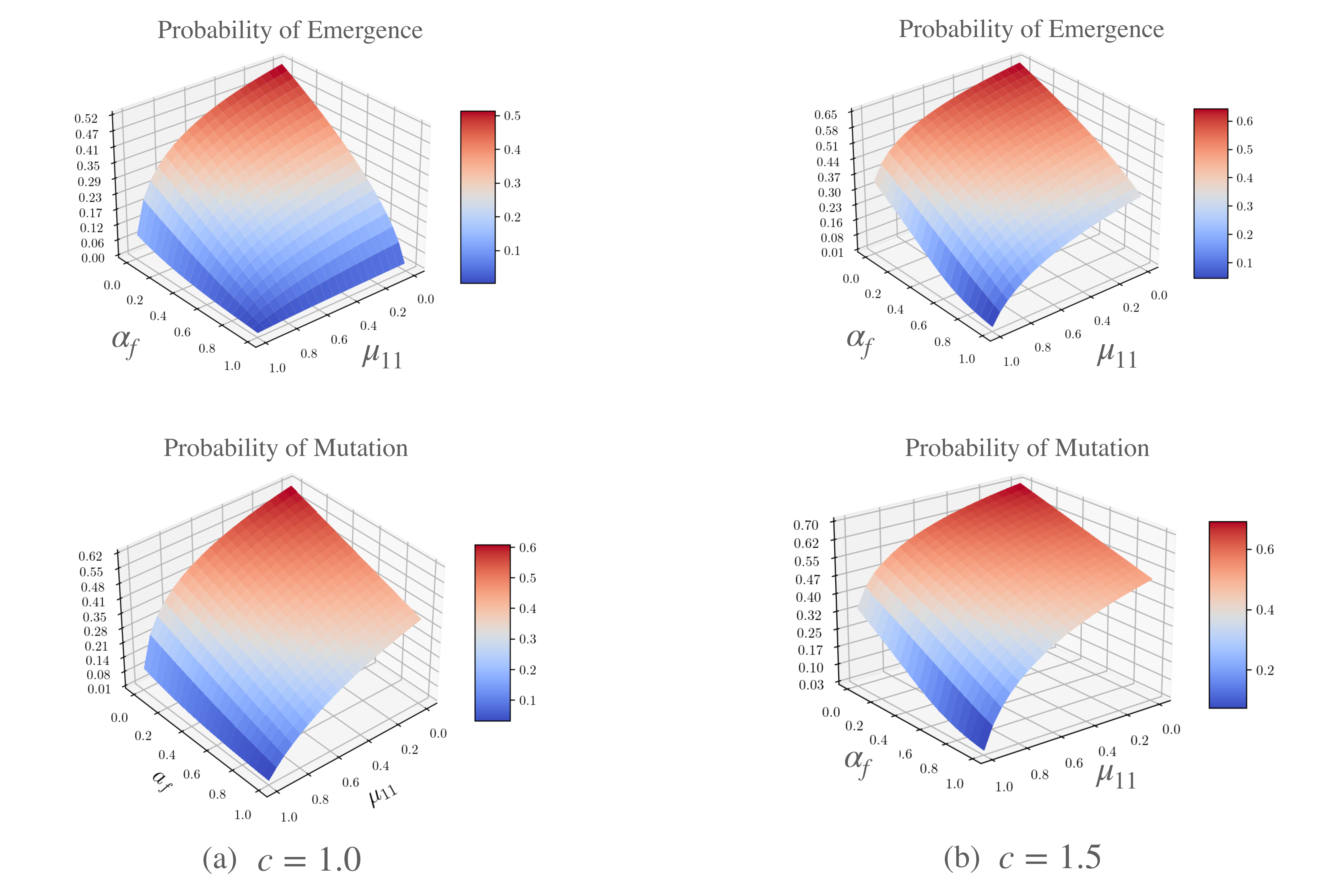}
		\caption{\sl For the case of one-step irreversible mutations \eqref{eq:onesteprev}, we plot the analytical i)probability of emergence of an epidemic (Theorem 1) and the ii) probability that a mutation to strain-2 appears in the chain of infections starting from strain-1 (denotes as the probability of mutation) (Lemma 3). We vary $(\alpha_f, \muoneone ) \in [0,1)\times[0,1)$ as characterized in Figure~\ref{fig:ne:surf-domain}. A smaller value of $\alpha_f$ and $\muoneone$ respectively correspond to a case with a greater participation in layer-$f$ and a higher probability of mutation to strain-2. We set $\tfone=0.4, \tftwo=0.8,  c= \twone/\tfone=\twtwo/\tftwo=1.0,1.5$, and $\nu_f=\nu_w=1.2$. We observe a low (less than $50\%$) probability of emergence when i) there is a single-layer present, i.e., $\alpha_f\rightarrow1$ (even with the highly contagious strain circulating), and ii) with a single strain circulating, i.e., $\muoneone\rightarrow1$ (even when the $f$ layer is added). However, when both layers are open and mutations occur with a positive probability, we see a higher probability of emergence.} 
		\label{fig:ne:surf-prob}
	\end{figure*} 

Next, in Figure~\ref{fig:ne:surf-domain}, we characterize the transition from the single-layer to multi-layer setting. We note that $\mu_{11}$, approaches one, it corresponds to the case of the spread of pathogens without mutations. Thus, the deviation of $\mu_{11}$ away from one provides a way to characterize the departure from the case where no mutations take place. 
%where there are no mutations and as the magnitude of $\mu_{11}$ decreseases there 
% the fitness landscape consists of two strains such that the emerg
% Emergence of a highly transmissible strain that may alter the course of the epidemic- the second layer provides more opportunity for transmission (and therefore mutations)
Similarly, to characterize the transition from a single-layer to a multi-layer network, we consider the following degree distributions for the two layers:
		\begin{align}
			&d^f\sim	\begin{cases}
				0 ~~\text{w.p.~~} \alpha_f
				\\
				{\rm Poisson} (\nu_f) ~~\text{w.p.~~} 1-\alpha_f,
			\end{cases}\nonumber\\
			& d^w \sim
				{\rm{Poisson}} (\nu_w). \label{eq:mix-poisson2}
		\end{align}
%In \eqref{eq:mix-poisson2} above, as $\alpha_f$ approaches one, no nodes participate in layer-$f$, while $\alpha_f=0$ corresponds to the scenario where we have two independent Poisson layers.
%by varying $\alpha_f \in [0,1)$ and the transition from single-strain to multi-strain with $\mu_{11} \in [0,1)$. 
As $\alpha_f \rightarrow 1$, no nodes participate in layer-$f$ and the contact network $\mathbb{H}$ only comprises a single-layer $\mathbb{W}$. While, as $\muoneone\rightarrow1$, no mutations to strain-2 appear with whp starting from strain-1. From Lemma 2, we know the epidemic threshold does not vary with $\muoneone$ and it is evident that the vertical line in Figure~\ref{fig:ne:surf-domain} corresponds to isocontour for epidemic threshold. 
% Note that at the corners of the domain $(\alpha_f, \muoneone ) \in [0,1]\times[0,1]$, the multi-strain multi-layer model effectively reduces to a simpler model as follows
% \begin{itemize}[noitemsep, leftmargin=*]
% \item $\alpha_f=1,\muoneone=0$: one layer $(\mathbb{W})$ and mutations to strain-2 occur whp (so effectively one strain),
% \item $\alpha_f=0,\muoneone=1$: two layers and there are no mutations to strain-2 whp,
% \item $\alpha_f=1,\muoneone=1$: one layer $(\mathbb{W})$ and there are no mutations to strain-2 whp (with high probability),
% \end{itemize}
% while all points in the interior of the domain $(\alpha_f, \muoneone ) \in [0,1]\times[0,1]$ and the point $\alpha_f=0,\muoneone=1$ correspond to the case where both types of edges and strains occur with a positive probability. 
%{\color{blue} Probability of emergence}

In Figure~\ref{fig:ne:surf-prob}, we investigate the joint impact of layer openings and mutation parameters on the probability of emergence of an epidemic outbreak. For the domain $(\alpha_f, \muoneone ) \in [0,1)\times[0,1)$ as described in Figure~\ref{fig:ne:surf-domain}, we plot the i) probability of emergence of an epidemic and the ii) probability that at least one mutation to strain-2 emerges in the chain of infections starting from strain-1, as given by Theorem 1 and Lemma 3 respectively. We set  $\tfone=0.4, \tftwo=0.8$, $c=10,1.2$ and 
$\nu_f=\nu_w=1.2$ in \eqref{eq:mix-poisson2}. In light of Lemma 2, since the epidemic threshold is not affected by $\muoneone$, it maybe tempting to consider a single-strain model as being sufficient to capture the epidemic characteristics. However, Figure~\ref{fig:ne:surf-prob} demonstrates that the possibility of mutations does matter in determining the likelihood of emergence of an epidemic. Moreover, we observe that for regions where there is effectively just a single-layer $(\alpha_f\rightarrow1)$, the probability of emergence remains low despite the possible emergence of a highly contagious strain.  Likewise, in cases where only a single, moderately-transmissible strain circulates with high probability  ($\muoneone\rightarrow1$), even the addition of another layer-$f$ does not lead to a high probability of emergence. In contrast, when both types of network layers are present, and there is a non-zero probability of mutation to strain-2, the probability of emergence is high. Moreover, there is a spectrum of intermediate values that the probability of emergence admits across the domain, with the single-layer or single-strain cases only capturing the limiting cases where $\alpha_f$ and $\muoneone$ respectively approach one. This observation sheds light on the impact of imposing/lifting mitigation measures concerning different contact network layers (e.g., school closures or many companies adopting work-from-home policies) on the emergence of more transmissible variants. For example, opening a new layer in the contact network may be deemed safe based on the transmissibility of the initial strain, but even a modest increase in infections caused by the new layer might increase the chances of a more transmissible strain to emerge, which in turn can make an epidemic more likely. Thus, by studying the mutations over a multi-layer network, our results can help understand the comprehensive impact of layer closures/openings.
		\subsection*{Reductions to simpler models}
		In this section we propose and analyze approaches to reduce the multi-strain multi-layer (MS-ML) model into either a single-layer or a single-strain. Next, we propose and evaluate different transformations for finding a corresponding single-strain multi-layer (SS-ML) or a multi-strain single-layer (MS-SL) model for a given MS-ML model.

		We first investigate the question that whether under arbitrary distributions for the network layers, can we systematically reduce the MS-ML model into an equivalent (yet simpler) MS-SL model \cite{ne:alexday} and get accurate predictions for key epidemiological quantities. In what follows, we show that when the degree distribution for the two network layers is Poisson with parameters $\lambda_f$ and $\lambda_w$ respectively, the following transformation to a MS-SL model can accurately predict the probability of emergence of an epidemic (\emph{\color{blue}SI Appendix 3.A}). 
		\begin{align}
\lambda &\leftarrow \lambda_f +  \lambda_w. \label{eq:mapto1layerlambda}\\
T_1 &\leftarrow \frac{\lambda_f  \tfone+ \lambda_w \twone}{\lambda_f + \lambda_w};~~ T_2 \leftarrow \frac{\lambda_f  \tftwo+ \lambda_w \twtwo}{\lambda_f + \lambda_w}. \label{eq:mapto1layer} 
	\end{align}

	We now illustrate the potential pitfall of mapping a multi-layer network to a single-layer structure using the transformations \eqref{eq:mapto1layerlambda} - \eqref{eq:mapto1layer} when the Poisson assumption no longer holds. As a concrete example %illustrating predictions made by models that coalesce the multi-layer structure into a single-layer, 
we study how well can the reduction to a single-layer allow us to predict the epidemic threshold for a multi-layer network for a more general family of distributions. We first consider the case when the ratio of the transmissibilities in the two network layers is one for both strains, i.e., $\frac{\twone}{\tfone} =\frac{\twtwo}{\tftwo}=1.$  In other words, the transmissibilities only depend on the type of strain and are agnostic of the type of link. Throughout, we let $\sigma(.)$ denote the epidemic threshold of the matrix supplied as its argument. Let the single-layer network obtained by taking the sum of node degrees in the two layers be denoted by $\tilde{\mathbb{H}}$. Since  the layers in the multi-layer network have independent degree  distributions, the degree distribution for $\tilde{\mathbb{H}}$ is given by $\tilde{p}=p^f * p^w$, where $*$  denotes the convolution operator. For $a \in \{f,w\}$, recall that $\lambda_a$ and $\beta_a$ respectively denote the mean degree distribution and the mean excess degree of distribution for network layer $a$.  It can be verified that for network $\tilde{\mathbb{H}}$, the mean degree distribution and mean excess degree distribution, respectively denoted by $\tilde{\lambda}$ and $\tilde{\beta}$ are given as:
		\begin{align}
			 \tilde{\lambda} &= \lambda_w +\lambda_f, \label{eq:mapping-lambda} \\
			 \tilde{\beta} &=  \frac{\beta_f \lambda_f + \beta_w \lambda_w +2 \lambda_f \lambda_w}{\lambda_f+\lambda_w}.  \label{eq:mapping-beta} 
		\end{align}
		%or equivalently the convolution of the pmf for the degree distribution for the two layers.
% 		\emph{Impact on the spectral radius:} 
From \eqref{eq:spec-prod}, it follows that for the multi-layer network  $\mathbb{H}=\mathbb{W} \cup \mathbb{F}$ , the critical threshold for the emergence  of the epidemic outbreak% in using \eqref{eq:spec-prod}, the spectral radius ($\rho_H$)
		   ~is% given as
		\begin{align}
			\sigma\left( \left[   \begin{matrix}
				\beta_f &  \lambda_w\\\ 
				\lambda_f &   \beta_w
			\end{matrix} \right] \right) \times \sigma\left(\pmb{T^f} \pmb{\mu}\right) >1.\label{eq:pred-thresh-2}
		\end{align}
Upon mapping the transmissibilities as per \eqref{eq:mapto1layer}, the critical threshold for the emergence  of the epidemic outbreak in $\tilde{\mathbb{H}}$ (constructed using the configuration model with degree drawn from the  distribution $p^f  * p^w$) is given by
\begin{align}
	%\tilde{\beta} &=
	  \frac{\beta_f \lambda_f + \beta_w \lambda_w +2 \lambda_f \lambda_w}{\lambda_f+\lambda_w} \times   \sigma\left(\pmb{T^f} \pmb{\mu}\right)  >1. \label{eq:pred-thresh-1}
\end{align}
Comparing the above thresholds in (\ref{eq:pred-thresh-2}) and (\ref{eq:pred-thresh-1}), we see that the predicted thresholds are identical if and only if
\begin{align}
				\sigma \left( \left[   \begin{matrix}
		\beta_f &  \lambda_w\\\ 
		\lambda_f &   \beta_w
	\end{matrix} \right] \right) =  \frac{\beta_f \lambda_f + \beta_w \lambda_w +2 \lambda_f \lambda_w}{\lambda_f+\lambda_w}  \label{eq:pred-thresh-1eq2}.
\end{align}
It can be verified that with $\lambda_f, \lambda_w, \beta_f, \beta_w>0$,  \eqref{eq:pred-thresh-1eq2} holds if and only if 
\begin{align}
	\beta_f - \lambda_f =\beta_w -\lambda_w \label{eq:matching-condition}.
\end{align}
We provide a proof of \eqref{eq:matching-condition} being equivalent to the condition \eqref{eq:pred-thresh-1eq2} in \emph{\color{blue}SI Appendix 3.B}.
The condition \eqref{eq:matching-condition} is equivalent to the \emph{dispersion indices} of the constituent network layers being the same, where the dispersion index is defined as the ratio of the variance and mean of a degree distribution. For cases when the degree distribution of the constituent network layers are not from the same parametric family of distributions, depending on the magnitude of the dispersion index relative to one, the condition \eqref{eq:matching-condition} may not hold. For instance, if the degree distribution of the two layers is respectively Poisson and Binomial, regardless of the choice of parameters of the distributions, the condition $\beta_f - \lambda_f =\beta_w -\lambda_w$ will never hold since the dispersion index of the Poisson distribution and the Binomial distribution are respectively $=1$ and $>1$. Likewise for the parameterization of the degree that accounts for non-participation in layer-$f$ \eqref{eq:mix-poisson2}, as long as $\alpha>0$, the dispersion indices of layers $f$ and $w$ do not match and thus reductions to MS-SL models can yield inaccurate predictions for the epidemic threshold.

So far we have investigated reductions to a multi-strain single-layer model (MS-SL). %looked at ..., good correspondence between MS-SL and MS-ML , 
Then, a natural question to ask is whether we can alternatively use a reduction to a single-strain multi-layer (SS-ML) model to characterize the epidemic? It is known \cite{ne:pnas,kenah2007second} that when there are correlations between infection events, the predictions made by models that assume independent transmission events can lead to incorrect predictions. For the multi-strain transmission model, the infection events are conditionally independent given the type of strain carried by a node and dependent otherwise. Therefore, models that do not account for correlation in infection events, such as single-strain spreading on multi-layer networks \cite{ne:conjoining},  can lead to inaccurate predictions for multi-strain settings; see \cite{ne:pnas} for a detailed discussion. 

Next, evaluate how transformations to SS-ML models predict the epidemic characteristics in presence of mutations. For reductions to SS-ML models, assuming $\frac{\twone}{\tfone} =\frac{\twtwo}{\tftwo}=c$, we consider the following two approaches. The first approach involves using \eqref{eq:spec-prod} and defining the equivalent transmissibilities for the two layers as:
\begin{align}
    &\rho(\pmb{T^f}\pmb{\mu}) \rightarrow \tilde{T}^f, \nonumber\\
        &\rho(\pmb{T^w}\pmb{\mu}) \rightarrow \tilde{T}^w (\equiv c \rho(\pmb{T^f}\pmb{\mu}) \rightarrow \Tilde{T}^w). \label{eq:ss-tx-rho}
\end{align}
Through \eqref{eq:spec-prod}, the transformation \eqref{eq:ss-tx-rho} ensures that the \emph{spectral radius} predicted by the corresponding SS-ML reduction is the same as the MS-ML model. 
The second approach is based on transforming to a single layer by directly matching the \emph{mean matrix}, i.e., we ensure that the mean number of secondary infections stemming from any given type of infected individual is the same across the models (\emph{\color{blue}SI Appendix 4}).

\begin{figure*}[t]
\centering 
\includegraphics[scale=0.32]{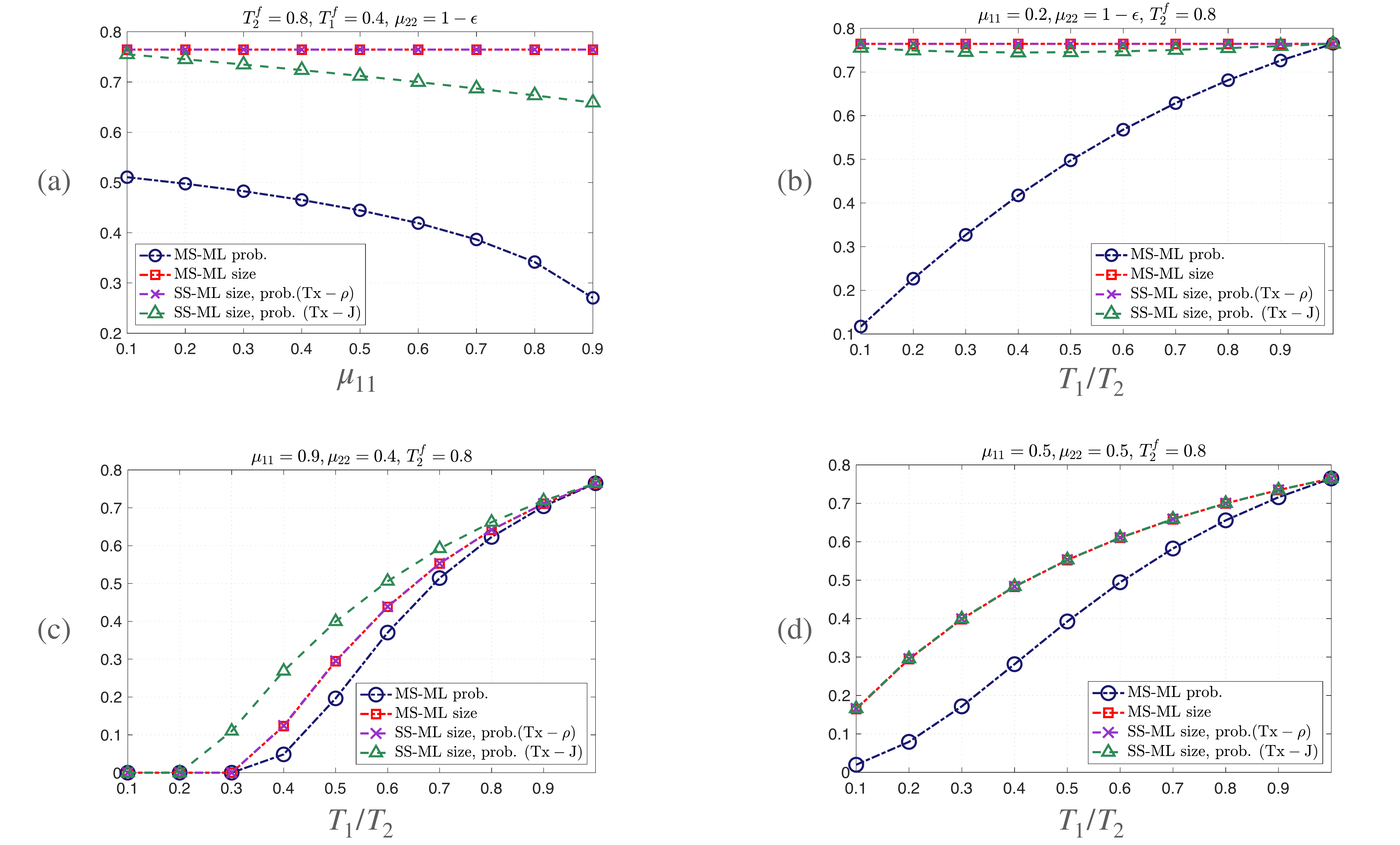}
\caption{\sl A comparison of the predictions made by reduction to a single-strain (SS-ML) through: mapping the spectral radius through (denoted as ${\rm Tx}-\rho$) and mapping the mean matrix for infections  (denoted as ${\rm Tx}-J$). (a) We set $\mutwotwo=1-\epsilon$ with $\epsilon=10^{-10}$ and vary $\muoneone$. (b) Throughout, we set $T^f_2=0.8$ and vary the ratio $T^f_1/T^f_2=T^w_1/T^w_2$, denoted as $T_1/T_2$ on the X-axis. We observe that neither transformation accurately predicts the probability of emergence and the gap in prediction is more pronounced when the difference in strain transmissibilities is higher, i.e., when $T_1/T_2$ is smaller. We observe that while the SS-ML reduction through mapping the spectral radius (${\rm Tx}-\rho$) captures the size but it fails to predict the probability of emergence.} \label{fig:ne:ss-ml-mu-prob}		  
\end{figure*}

 Next, we evaluate the above reductions to SS-ML models.
 Following the network distribution in \eqref{eq:mix-poisson2}, we set $\alpha_f = 0.2$,
$\alpha_w=0$,~$\nu_f=\nu_w=1.2$, $\tftwo=0.8$, and we assume ${\twone}/{\tfone} ={\twtwo}/{\tftwo}=c=1.2$. In Figure~\ref{fig:ne:ss-ml-mu-prob}(a), we fix the transmissibility parameters and plot the predictions made by reduction to a single-strain (SS-ML) through mapping the spectral radius (denoted as ${\rm Tx}-\rho$) and mapping the mean matrix for infections (denoted as ${\rm Tx}-J$). We set $\mutwotwo=1-\epsilon$ with $\epsilon=10^{-10}$ and vary $\muoneone$ in the interval $[0,1]$. From Lemma 2, we know that as $\mutwotwo \rightarrow 1$, $\rho(\pmb{T^f}\pmb{\mu})$ remains constant, and thus, the prediction made by the SS-ML under the spectral radius mapping remains constant. We note that SS-ML models coalesce the transmissibility of the two strains into effective transmissibility for each layer and predict the same value of the probability of emergence and the epidemic size given. Substituting $\tfone=\tftwo=\rho(\pmb{T^f}\pmb{\mu})$ and $\twone=\twtwo=c\rho(\pmb{T^f}\pmb{\mu})$ and $\pmb{\mu^f}=\pmb{\mu^w}=\pmb{I_2}$ in Theorem 1, we get the predictions for the SS-ML model under the ${\rm Tx}-\rho$. We note that that while the SS-ML reduction through mapping the spectral radius (${\rm Tx}-\rho$) captures the size but it fails to predict the probability of emergence. This observation is consistent with the observation that reducing an MS-SL model to a bond-percolation model leads to inaccurate prediction for the probability of emergence but correctly predicts the total epidemic size \cite{ne:pnas}.

In Figures~\ref{fig:ne:ss-ml-mu-prob}(b)-(d), we plot how the predictions of the SS-ML model are impacted as a function of how different the transmissibilities for the two strains for different values of the mutation parameters. We do so by by varying the ratio $\tfone/\tftwo=\twone/\twtwo$, denoting it by $T_{1}/T_{2}$ while keeping $\tftwo$ constant at 0.8. We also observe that mapping through ${\rm Tx}-J$ neither accurately predicts the epidemic size nor the probability of emergence. We observe that the gap in the prediction of the probability of emergence by the SS-ML models is less pronounced when $T_1/T_2 \rightarrow 1$. Further, when $\mutwotwo \rightarrow 1$, the predictions made by mapping the spectral radius are constant in line with Lemma 2. Whereas, when $\mutwotwo=0.5,0.9$, we see that the SS-ML size predictions under the ${\rm Tx}-\rho$ transformation vary with the ratio $T_1/T_2$. As above, the SS-ML model under the ${\rm Tx}-\rho$ transformation captures the size and the epidemic threshold while failing to predict the probability of emergence accurately. We observe that when $\muoneone=\mutwotwo=0.5$, the predictions made by both the SS-ML transforms align with the total epidemic size obtained with the MS-ML model. One general shortcoming of the SS-ML transformations is that they predict the same probability of emergence and epidemic size; they can at best only predict one of these metrics accurately. Moreover, they do not shed light on the fraction of individuals infected by each strain type. Also, we note that both the transformations (matching the epidemic threshold and the mean mutation matrix) critically relied on the decoupling \eqref{eq:spec-prod}, which only holds when $\tfone/\twone = \tftwo/\twtwo$. Our observations further highlight the importance of developing epidemiological models that account for the heterogeneity in network structure and pathogen strains. 

	\subsection*{Data Availability}
	Data will be available upon request from the corresponding author.
 \subsection*{Conclusions}
 This work analyzed the spreading characteristics of mutating contagions over multi-layer networks. We derived the fundamental epidemiological quantities for the proposed multi-layer multi-strain model: the probability of emergence, the epidemic threshold, and the mean fraction of individuals infected with each strain. Our results highlight that the impact of imposing/lifting mitigation measures concerning different contact network layers should be evaluated in connection with their effect on the emergence of a new pathogen strain. Furthermore, we proposed and analyzed transformations to simpler models that do not simultaneously account for the heterogeneity in pathogen strains and network layers. We showed that existing models cannot be invoked to accurately characterize the multi-layer multi-strain setting while also unraveling conditions under which simpler models can be helpful for making predictions. An important future direction is to extend the network-based analysis of multi-strain spreading \cite{ne:alexday} to allow for the case when the previous infection with one strain confers full immunity only with respect to that strain while leaving the individual susceptible to other strains, although possibly at a reduced level. While \emph{cross-immunity} among different pathogen strains has been studied \cite{kryazhimskiy2007state} in multi-spreading models that do not account for the contact network structure, it is of interest to research cross-immunity interference in the light of different contact patterns.  Such an analysis will pave the way for evaluating the risk of the emergence of new strains that can evade immunity acquired from previous infections or vaccination in light of different policy measures that alter the contact network. A promising future application is to leverage models for multi-strain spreading to combat the spread of misinformation across different social media platforms.
		\subsection*{Acknowledgements}
	{This research was supported in part by: the National Science Foundation through grants 2225513, 2026985, 1813637, CCF-2142997, CNS-2041952, and CCF-1917819; the Army Research Office through grants  W911NF-22-1-0181, W911NF-20-1-0204 and W911NF-18-1-0325; an IBM academic award; a gift from Google; and Dowd Fellowship, Knight Fellowship, Lee-Stanziale Ohana Endowed Fellowship, and Cylab Presidential Fellowship from Carnegie Mellon University.}
	
		\newpage

	\section*{References}
	% Bibliography
	\bibliography{pnas-sample}
	\newpage
 \onecolumn
 \section*{SI Appendix}
 \setcounter{section}{0}
 \section{Materials and Methods}
\label{sec:prob}
\subsection{Proof of Theorem 1 (Probability of Emergence)} {\color{white}$-$}\\
\emph{Preliminaries:}
The proof of Theorem~1 uses the fact that the network $\mathbb{H}$ is locally tree-like, which follows from the result \cite{Soderberg3} that the clustering coefficient of colored degree-driven networks scales as $1/n$ as $n$ gets large. The key idea behind this proof is based on the bond percolation analysis for the configuration model \cite[p.~385]{ne:newman2018networks} where it is argued that a node does not belong to the giant component if and only if it is not connected to the giant component via any of its neighbors. This observation is extended to the multi-strain setting by noting that a randomly selected node $v$ infects only a finite number of nodes if and only if all neigbors of $v$ infect only a finite number of nodes \cite{ne:alexday}. \\
% %
%     \cite{ne:newman2018networks}
%     [Page 385, Networks, Newman 2018]
%     the key idea behind this approach for the configuration model without percolation- \\Alexander and Day We use the observation: 
\noindent\emph{Derivation of PGFs:} For completing the proof of Theorem~1, we derive the PGFs for infected neighbors of the seed node and later generation infectives. Recall that the seed node is infected with strain-1. For $i \in \{1,2\}$ and $a \in \{f,w\}$, let $X^a_i$ denote the number of infections of type-$i$ transmitted by a seed node $u$ to its neighbors which are connected through type-$a$ edges. We first derive the PGF $\gamma_1(z^f_1,z^f_2, z^w_1,z^w_2 )$ for $X^f_1,X^f_2,X^w_1,X^w_2 $ i.e., the number of infection events of each type induced among the neighbors of a seed node when the seed node is infected with strain-1. 

\begin{align}
	&\gamma_1(z^f_1,z^f_2, z^w_1,z^w_2 )\nonumber\\
	&= \E \left[(z^f_1)^{X^f_1}(z^f_2)^{X^f_2} (z^w_1)^{X^w_1}(z^w_2)^{X^w_2}  \right] \nonumber\\ 
	&= \E \left[\E\left[
	(z^f_1)^{X^f_1}(z^f_2)^{X^f_2} (z^w_1)^{X^w_1}(z^w_2)^{X^w_2} ~|~ \boldsymbol{d}_u \right]\right]\nonumber \\ 
	&= \E \Bigg [\left(\sum_{x^f_1=0}^{d^f_u}\sum_{x^f_2=0}^{d^f_u-x^f_1} \binom{d^f_u}{x^f_1}\binom{d^f_u-x^f_1}{x^f_2}(1-\tfone)^{d^f_u-x^f_1-x^f_2} (\tfone \mufoneone)^{x^f_1} (\tfone \mufonetwo)^{x^f_2} 
	(z^f_1)^{x^f_1}(z^f_2)^{x^f_2}\right)\nonumber\\
	&  \qquad \cdot
	\left(\sum_{x^w_1=0}^{d^w_u}\sum_{x^w_2=0}^{d^w_u-x^w_1} \binom{d^w_u}{x^w_1}\binom{d^w_u-x^w_1}{x^f_2}
	(1-\twone)^{d^w_u-x^w_1-x^w_2} (\twone \muwoneone)^{x^w_1} (\twone \muwonetwo)^{x^w_2} (z^w_1)^{x^w_1}(z^w_2)^{x^w_2} \right)\Bigg ] \nonumber%~\Bigg|~ \boldsymbol{d}_u 
	\\ 
	&=\E \Big[  (1-\tfone +\tfone \mufoneone z^f_1 +\tfone \mufonetwo z^f_2 )^{d^f_u}  (1-\twone +\twone \muwoneone z^w_1 +\twone \muwonetwo z^w_2 )^{d^f_u}\Big] \nonumber\\
	&=\sum_{\boldsymbol{d}} p_{\boldsymbol{d}} (1-\tfone +\tfone \mufoneone z^f_1 +\tfone \mufonetwo z^f_2 )^{d^f_u}  (1-\twone +\twone \muwoneone z^w_1 +\twone \muwonetwo z^w_2 )^{d^f_u} \nonumber \\
	&= g \left(1-\tfone+\tfone\left( \sum_{j=1}^2 \mu^f_{1j} z^f_j \right), 1-\twone+\twone \left(\sum_{j=1}^2 \mu^w_{1j} z^w_j \right)\right)\nonumber.
\end{align}
Following the above sequence of steps, it is easy to verify that for $a \in \{f,w\}$ and $i \in \{1,2\}$, $\Gamma^a_i(z^f_1,z^f_2, z^w_1,z^w_2)$ is the PGF for the number of infection events of each type caused by a {\em later-generation} infective that acquired the infection through a type-$a$ edge and carries strain-$i$ after mutation. In particluar, 
\begin{align}
	\Gamma^f_1(z^f_1,z^f_2, z^w_1,z^w_2 )&= G^f \left(1-T^f_1+T^f_1\left( \sum_{j=1}^2 \mu^f_{1j} z^f_j \right), 1-T^w_1+T^w_1 \left(\sum_{j=1}^2 \mu^w_{1j} z^w_j \right) \right) \nonumber \\
	\Gamma^f_2(z^f_1,z^f_2, z^w_1,z^w_2 )&= G^f \left(1-T^f_2+T^f_2 \left(\sum_{j=1}^2 \mu^f_{2j} z^f_j\right), 1-T^w_2+T^w_2 \left(\sum_{j=1}^2 \mu^w_{2j} z^w_j \right) \right)\nonumber \\
	\Gamma^w_1(z^f_1,z^f_2, z^w_1,z^w_2 )&= G^w \left(1-T^f_1+T^f_1 \left(\sum_{j=1}^2 \mu^f_{1j} z^f_j\right), 1-T^w_1+T^w_1\left( \sum_{j=1}^2 \mu^w_{1j} z^w_j \right)\right)  \nonumber \\
	\Gamma^w_2(z^f_1,z^f_2, z^w_1,z^w_2 )&= G^w \left(1-T^f_2+T^f_2 \left(\sum_{j=1}^2 \mu^f_{2j} z^f_j\right), 1-T^w_2+T^w_2 \left(\sum_{j=1}^2 \mu^w_{2j} z^w_j \right)\right).\nonumber 
\end{align}
\emph{Proof}:
For computing the probability of emergence, we define its complementary event-- the probability of extinction, i.e, the probability that the disease outbreaks infects only a finite number of individuals and eventually dies out. We express the probability of extinction starting from an initial infective (denoted by $\pr {[\rm Extinction]}$) in terms of the probability of extinction starting from a later generation infective.  For $a \in \{w,f\}$ and $i \in \{1,2\}$, let $q^a_i$ denote the probability of extinction starting from one later-generation infective carrying strain-$i$ which was infected through a type-$a$ edge. Using the observation that the seed node infects only a finite number of nodes if and only if all its neigbors infect only a finite number of nodes \cite{ne:alexday}, we get the following recursions between the probability of extinction of the epidemic starting from the seed node in terms of the the probability of extinction starting from later-generation infectives $(q^f_1,q^f_2, q^w_1,q^w_2)$. We have,

%, denoted $q^a_i$, is the smallest non-negative root of the equation $q^a_i = \Gamma^a_i \left(q^f_1,q^f_2, q^w_1,q^w_2\right)$ solved simultaneously for $i=1,2$. 

\begin{align}
	&	\pr {[\rm Extinction]}\nonumber\\
	&= \sum_{x^f_1,x^f_2,x^w_1,x^w_1}
	\pr {{[\rm Extinction}~|~X^f_1=x^f_1,X^f_2=x^f_2,X^w_1=x^w_1,X^w_2=x^w_2 }] \pr[X^f_1=x^f_1,X^f_2=x^f_2,X^w_1=x^w_1,X^w_2=x^w_2 ] \nonumber\\
	&= \sum_{x^f_1,x^f_2,x^w_1,x^w_1}
	(q^f_1 )^{x^f_1} (q^f_2 )^{x^f_2 } (q^w_1 )^{x^w_1} (q^w_2 )^{x^w_2}	\pr[X^f_1=x^f_1,X^f_2=x^f_2,X^w_1=x^w_1,X^w_2=x^w_2 ]\nonumber\\
	&=\gammaoneQ.\label{eq:ne:extinction-expr}
\end{align}
Similarly, for the later-generation infectives, we have
\begin{align}
	\qfone&=\GammafoneQ\nonumber\\
	\qftwo&=\GammaftwoQ \nonumber\\
	\qwone&=\GammawoneQ\nonumber\\
	\qwtwo&=\GammawtwoQ \nonumber.
\end{align}
Note that $q^f_1=q^f_2=q^w_1=q^w_2=1$ is a trivial solution of the above fixed point equations. We derive the smallest non-negative root of the above fixed point equations and substitute in (\ref{eq:ne:extinction-expr}) to obtain the desired result:
\begin{align}
	%Probability of extinction  
	\pr {[\rm Emergence]} =1- \gammaoneQ.
\end{align}
\hfill{$\blacksquare$}
\subsection{Proof of Theorem~2 (Epidemic Threshold)}
\emph{Deriving the Jacobian matrix}: The epidemic threshold is ascertained by determining the stability of the fixed point of the recursive equations in Theorem~2 by linearization around $\qfone=\qftwo=\qwone=\qwtwo=1$ which yields the Jacobian matrix
in Theorem~2 as below.
%\subsection{Multi-type branching processes}

\begin{align}
	\pmb{J}&= \left[
	\begin{matrix}
		\frac{\partial \GammafoneQ}{\partial q^f_1} & \frac{\partial\GammafoneQ}{\partial q^f_2}  & \frac{\partial\GammafoneQ}{\partial q^w_1} & \frac{\partial\GammafoneQ}{\partial q^w_2}\\ 
		\frac{\partial \GammaftwoQ}{\partial q^f_1} & \frac{\partial\GammaftwoQ}{\partial q^f_2}  & \frac{\partial\GammaftwoQ}{\partial q^w_1} & \frac{\partial\GammaftwoQ}{\partial q^w_2}\\ 
		\frac{\partial \GammawoneQ}{\partial q^f_1} & \frac{\partial\GammawoneQ}{\partial q^f_2}  & \frac{\partial\GammawoneQ}{\partial q^w_1} & \frac{\partial\GammawoneQ}{\partial q^w_2}\\ 
		\frac{\partial \GammawtwoQ}{\partial q^f_1} & \frac{\partial\GammawtwoQ}{\partial q^f_2}  & \frac{\partial\GammawtwoQ}{\partial q^w_1} & \frac{\partial\GammawtwoQ}{\partial q^w_2}
	\end{matrix} \right]_{q^f_1,q^f_2, q^w_1,q^w_2=1} \label{eq:ne:Jmatrixdef}.
\end{align}
For $a,b \in \{f,w\}$ and $i,j \in \{1,2 \}$, we have
\begin{align} 
	\Gamma^a_i(z^f_1,z^f_2, z^w_1,z^w_2 )&= %G^a \left(1-T^f_i+T^f_i  \mu^f_{i1} q^f_1+T^f_i  \mu^f_{i2} q^f_2, 1-T^w_i+T^w_i  \mu^w_{i1} q^w_1+T^w_i  \mu^w_{i2} q^w_2 \right), 
	G^a \left(1-T^f_i+T^f_i \left(\sum_{j=1}^2 \mu^f_{ij} z^f_j \right), 1-T^w_i+T^w_i \left(\sum_{j=1}^2 \mu^w_{ij} z^w_j \right)\right) \nonumber\\
	\frac{\partial 	\Gamma^a_i(q^f_1,q^f_2, q^w_1,q^w_2 )}{\partial q^b_j}	\Biggr|_{\substack{q^f_1,q^f_2, q^w_1,q^w_2=1}} &=  \begin{cases}
		\frac{\partial G^a(z^f,z^w)}{\partial z^f} \Bigr|_{\substack{z^f,z^w=1}}T_i^f \mu^f_{ij}, 
		& \text{  if } b=f \\
		\frac{\partial G^a(z^f,z^w)}{\partial z^w} \Bigr|_{\substack{z^f,z^w=1}}T_i^w \mu^w_{ij}, & \text{  if } b=w. \label{eq:ne:partial-cases}
	\end{cases}
	%\frac{\partial G^a(z^f,z^w)}{\partial z^f} \Bigr|_{\substack{z^f,z^w=1}}\frac{\partial  (1-T^f_i+T^f_i  \mu^f_{i1} q^f_1+T^f_i  \mu^f_{i2} q^f_2) }{\partial q^f_j}=
\end{align}
From the properties of PGFs, note that for a later-generation infective which acquired  a type-$i$ strain (after mutation) through a type-$a$ contact, equation (\ref{eq:ne:partial-cases}) gives the mean number of neighbors/offsprings which were infected with type-$j$ strain (after mutation) through type-$b$ contacts.
\begin{align}
	\frac{	\partial G^f(z^f,z^w)}{\partial z^f} \Bigr|_{\substack{z^f,z^w=1}}&=  \frac{ \langle d_f^2 \rangle  -\lambda_f }{\lambda_f}=\beta_f& \frac{	\partial G^f(z^f,z^w)}{\partial z^w}\Bigr|_{\substack{z^f,z^w=1}}= \lambda_w \label{eq:ne:partial-cases-f}\\
	\frac{	\partial G^w(z^f,z^w)}{\partial z^f} \Bigr|_{\substack{z^f,z^w=1}}&=\lambda_f  &\frac{	\partial G^w(z^f,z^w)}{\partial z^w}\Bigr|_{\substack{z^f,z^w=1}}= \frac{ \langle d_w^2 \rangle  -\lambda_w }{\lambda_w}=\beta_w.\label{eq:ne:partial-cases-w}
\end{align}
Substituing (\ref{eq:ne:partial-cases}), (\ref{eq:ne:partial-cases-f}) and (\ref{eq:ne:partial-cases-w}) in (\ref{eq:ne:Jmatrixdef}) yields
\begin{align}
	\boldsymbol{J}= \left[   \begin{matrix}
		\tfone \mufoneone \beta_f &  \tfone \mufonetwo\beta_f &\twone \muwoneone \lambda_w & \twone \muwonetwo \lambda_w\\ 
		\tftwo \muftwoone \beta_f &  \tftwo \muftwotwo\beta_f & \twtwo \muwtwoone \lambda_w & \twtwo \muwtwotwo  \lambda_w\\
		\tfone \mufoneone\lambda_f & \tfone \mufonetwo \lambda_f & \twone \muwoneone \beta_w & \twone \muwonetwo \beta_w\\ 
		\tftwo \muftwoone \lambda_f &   \tftwo \muftwotwo \lambda_f &\twtwo \muwtwoone\beta_w &\twtwo \muwtwotwo \beta_w 
	\end{matrix} \right].
\end{align}
%\subsection{Stability of fixed point solutions}
Note that $q^f_1=q^f_2=q^w_1=q^w_2=1$ is always a solution of the above fixed point equations.  An epidemic emerges starting from a seed with strain-1 when there is a positive probability that the pathogen escapes extinction in a later generation infective, i.e., when  $q^a_i <1$ for some $a \in \{f,w\}, i \in\{1,2\}$ for which $\mu^a_{1i}>0$. Invoking the theory from multi-type branching processes, we say that matrix $\pmb{J}$ corresponds to an indecomposable multi-type branching process if there is a positive probability that a node $u$ which received infection through a type-$a$ edge and acquired the type-$i$ strain after mutation, results in the event that a node $v$ receives the infection from a type-$b$ link and acquires the type-$j$ strain after mutation, after a finite number of steps for all $a,b \in \{f,w\}, i,j \in\{1,2\}$. %Note that the entry $\pmb{J}_{k,l}$ denotes 
When $\pmb{J}$ represents an indecomposable multi-type branching process, it is known \cite{ne:alexday} that extinction occurs with probability 1 from any later generation infective, i.e., $q^f_1=q^f_2=q^w_1=q^w_2=1$ is the smallest fixed point if and only if $\rho(\pmb{J})\leq1$; while if $\rho(\pmb{J})>1$, then $0\leq q^a_i <1$ for all $a \in \{f,w\}, i \in\{1,2\}$ and thus there is a positive probability that  the pathogen escapes extinction \cite{ne:alexday}. For decomposable processes, as long as there is no type that produces exactly one offpsring in its class with probability one, we still get that extinction occurs with probability 1 if and only if $\rho(\pmb{J})\leq1$ \cite{ne:alexday}. Therefore, the critical threshold for the emergence of the epidemic is $\rho(\pmb{J})=1$ and the super-critical regime corresponds to $\rho(\pmb{J})>1$.\hfill{$\blacksquare$}
\subsection{Proof of Theorem~3 (Mean Epidemic Size)}
% \emph{Definitions: $f_i(u^f_1,u^f_2,u^w_1,u^w_2, z^f, z^w)$}:
Throughout, we denote
	\begin{align}
		&f_i(u^f_1,u^f_2,u^w_1,u^w_2, z^f, z^w) :=&\nonumber \\
		&  \sum_{k^f_1=0}^{z^f} \sum_{k^f_2=0}^{z^f-k^f_1} \sum_{k^w_1=0}^{z^w} \sum_{k^w_2=0}^{z^w-k^w_1}\Bigg[
		&\binom{z^f}{k^f_1} \binom{z^f-k^f_1}{k^f_2} \left( u^f_1 \right)^{k^f_1}  \left( u^f_2 \right)^{k^f_2} \left( 1-u^f_1-u^f_2 \right)^{z^f-k^f_1-k^f_2} \nonumber  \\
		& &\cdot  \binom{z^w}{k^w_1} \binom{z^w-k^w_1}{k^w_2} \left( u^w_1 \right)^{k^w_1}  \left( u^w_2 \right)^{k^w_2} \left( 1-u^w_1-u^w_2 \right)^{z^w-k^w_1-k^w_2} \nonumber\\
		& &  \cdot \sum_{x^f_1=0}^{k^f_1} \sum_{x^f_2=0}^{k^f_2} \sum_{x^w_1=0}^{k^w_1} \sum_{x^w_2=0}^{k^w_2} \Bigg\{
		\binom{k^f_1}{x^f_1} \binom{k^f_2}{x^f_2} (T^f_1)^{x^f_1} (T^f_2)^{x^f_2} (1-T^f_1)^{k^f_1-x^f_1} (1-T^f_2)^{k^f_2-x^f_2}  \nonumber \\
		& & \cdot \binom{k^w_1}{x^w_1} \binom{k^w_2}{x^w_2} (T^w_1)^{x^w_1} (T^w_2)^{x^w_2}(1-T^w_1)^{k^w_1-{x^w_1}} (1-T^w_2)^{k^w_2-x^w_2}  \nonumber \\
		&  &  \cdot   \Bigg(  \frac{\mu^f_{1i} x^f_1+\mu^f_{2i}x^f_2+\mu^w_{1i}x^w_1+\mu^w_{2i}x^w_2}{x^f_1+x^f_2+x^w_1+x^w_2}\pmb{1}[x^f_1+x^f_2+x^w_1+x^w_2>0] \Bigg) \Bigg\}\Bigg]	\label{ne:eq:size-rec-f}.
	\end{align}
% 	For $i=1,2$, we have 
% 	\begin{align}
% 		q^f_{\ell+1, i} &= \sum_{\boldsymbol{d} } \frac{d^f  p_{\boldsymbol{d}}}{\langle d^f \rangle }  f_i(q^f_{\ell,1},q^f_{\ell,2},q^w_{\ell,1},q^w_{\ell,2},d^f-1, d^w),	\label{ne:eq:size-rec-levels-f}\\
% 		q^w_{\ell+1, i} &=  \sum_{\boldsymbol{d} } \frac{d^w  p_{\boldsymbol{d}}}{\langle d^w \rangle }  f_i(q^f_{\ell,1},q^f_{\ell,2},q^w_{\ell,1},q^w_{\ell,2},d^f, d^w-1)
% 		\label{ne:eq:size-rec-levels-w}
% 	\end{align}
\emph{Proof:}
Consider an arbitrary node $u$ at level $\ell+1$.	For $i \in \{1,2\}, a \in \{f,w\}$, let $Z^a$ denote the total number of type-$a$ contacts of node $u$ at level $\ell$ and let $K^a_i$ denote the number of active type-$a$ contacts of node $u$ that carry strain-$i$ at level $\ell$ . In the steps below, we express $q^f_{\ell+1, 1}$ in terms of $q^a_{\ell, i}$, where $i \in \{1,2\}, a \in \{f,w\}$. Recall that the degree distribution of node $u$ reached through a type-$f$ edge (from a node in level $\ell+2$) is given by $\frac{d^f  p_{\boldsymbol{d}}}{\langle d^f \rangle }$. Let $\mathcal{E}$ denote the event that $u$ gets infected with strain-$1$ through contact with neighbors in level $\ell$. We have%Let  $\mathcal{E}$ denote the event that node $u$ at level carries strain-$i$ (after mutation) 
\begin{align}
	q^f_{\ell+1, i}&= \sum_{d^f,d^w } \frac{d^f  p_{\boldsymbol{d}}}{\langle d^f \rangle }  \pr[{\mathcal{E}~|~  Z^f= d^f-1, Z^w=d^w}]\nonumber\\
	&=\sum_{d^f,d^w } \frac{d^f  p_{\boldsymbol{d}}}{\langle d^f \rangle } \Bigg(\sum_{k^w_1=0}^{d^w} \sum_{k^w_2=0}^{d^w-k^w_1} \sum_{k^f_1=0}^{d^f} \sum_{k^f_2=0}^{d^f-k^f_1}
	\pr[{\mathcal{E}~|~  Z^f= d^f-1, Z^w=d^w,K^f_1=k^f_1,K^f_2=k^f_2,K^w_1=k^w_1,K^w_2=k^w_2}]\nonumber\\
	&\qquad \qquad  \qquad \cdot \pr[K^f_1=k^f_1,K^f_2=k^f_2,K^w_1=k^w_1,K^w_2=k^w_2~|~Z^f= d^f-1, Z^w=d^w]
	\Bigg) \label{ne:eq:size-rec-proof-1}
\end{align}
Note that for $a\in\{f,w\}, i\in\{1,2\}$, given the random variable $Z^a$, we have $K^a_i={\rm Binomial}(Z^a,q^a_{\ell,i})$. Further, given $Z^f$ (resp., $Z^w$), the tuple $(K^f_1,K^f_2, Z^f-K^f_1-K^f_2)$ (resp., $K^w_1,K^w_2, Z^w-K^w_1-K^w_2$) follow independent multinomial distributions and therefore 
\begin{align}
	&	\pr[K^f_1=k^f_1,K^f_2=k^f_2,K^w_1=k^w_1,K^w_2=k^w_2~|~Z^f= d^f-1, Z^w=d^w]=\nonumber\\
	&\quad =\binom{z^f}{k^f_1} \binom{z^f-k^f_1}{k^f_2} \left( q^f_{\ell,1} \right)^{k^f_1}  \left( q^f_{\ell,2} \right)^{k^f_2} \left( 1-q^f_{\ell,1}-q^f_{\ell,2} \right)^{z^f-k_1-k_2}  \binom{z^w}{k^w_1} \binom{z^w-k^w_1}{k^w_2} \left( u^w_1 \right)^{k^w_1}  \left( u^w_2 \right)^{k^w_2} \left( 1-u^w_1-u^w_2 \right)^{z^w-k_1-k_2} \label{ne:eq:size-rec-proof-2}
\end{align}
For obtaining in summation (\ref{ne:eq:size-rec-f})
we further condition on the number of active contacts who successfully transmit the infection to node $u$. Suppose, $u$ receives $X^f_1$ (resp., $X^w_1$) infections of strain-$1$ through type-$f$ (resp., type-$w$) contacts, and  $X^f_2$ (resp., $X^w_2$) infections of strain-$2$ through type-$f$ (resp., type-$w$) contacts, we have
\begin{align}
	 & \pr[{\mathcal{E}~|~  Z^f= d^f-1, Z^w=d^w,K^f_1=k^f_1,K^f_2=k^f_2,K^w_1=k^w_1,K^w_2=k^w_2}] \nonumber                                                                                                                                      \\
	 & \quad=  \sum_{x^f_1=0}^{k^f_1} \sum_{x^f_2=0}^{k^f_2} \sum_{x^f_1=0}^{k^f_1} \sum_{x^f_2=0}^{k^f_2}
	\binom{k^f_1}{x^f_1} \binom{k^f_2}{x^f_2} (T^f_1)^{x^f_1} (T^f_2)^{x^f_2} (1-T^f_1)^{k^f_1-x^f_1} (1-T^f_2)^{k^f_2-x^f_2}  \nonumber \\
	 & \qquad \qquad  \qquad \qquad \qquad \cdot  \binom{k^w_1}{x^w_1} \binom{k^w_2}{x^w_2} (T^w_1)^{x^w_1} (T^w_2)^{x^w_2} (1-T^w_1)^{k^w_1-{x^w_1}} (1-T^w_2)^{k^w_2-{x^w_2}}  \nonumber                                                       \\
	 & \qquad \qquad  \qquad \qquad \qquad \cdot\Bigg(  \frac{\mu^f_{1i} 	 x^f_1+\mu^f_{2i}x^f_2+\mu^w_{1i}x^w_1+\mu^w_{2i}x^w_2}{x^f_1+x^f_2+x^w_1+x^w_2}\pmb{1}[x^f_1+x^f_2+x^w_1+x^w_2>0] \Bigg)\label{ne:eq:size-rec-proof-3}
\end{align}
Substituting (\ref{ne:eq:size-rec-proof-2}) and (\ref{ne:eq:size-rec-proof-3}) in (\ref{ne:eq:size-rec-proof-1}), we get for $i=1$,
	\begin{align}
		q^f_{\ell+1, i} &= \sum_{\boldsymbol{d} } \frac{d^f  p_{\boldsymbol{d}}}{\langle d^f \rangle }  f_1(q^f_{\ell,1},q^f_{\ell,2},q^w_{\ell,1},q^w_{\ell,2},d^f-1, d^w).\nonumber
	\end{align}
Similarly, we can obtain the recurrence equation for  
$q^f_{\ell+1, 2}$, $q^w_{\ell+1, 1}$, and $q^w_{\ell+1, 2}$. Using the limiting values $q^a_{\infty,i}$, for $a\in \{f,w\}, i \in \{1,2\}$, and noting that all the edges incident on the root node arise from the level below the root yields 		\begin{align}
		Q_i &= \sum_{{\boldsymbol{d}}}p_{\boldsymbol{d}} f_i(q^f_{\infty,1},q^f_{\infty,2},q^w_{\infty,1},q^w_{\infty,2},d^f, d^w). \nonumber
	\end{align}.\hfill{$\blacksquare$}
	
\subsection{Power law degree distribution with exponential cutoff}
			\begin{figure}[t]
		\centering
		\includegraphics[scale=0.29]{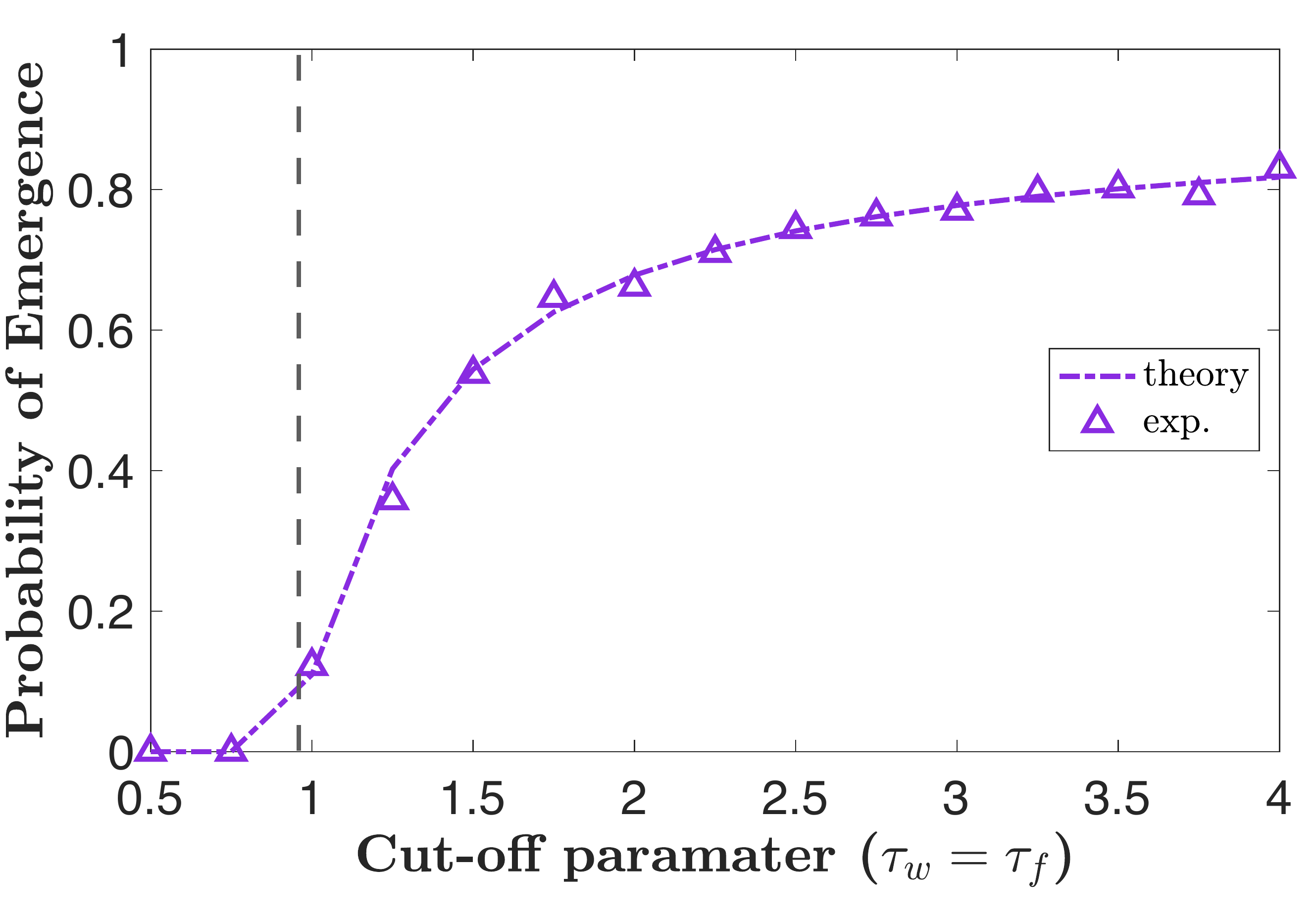}
				\includegraphics[scale=0.29]{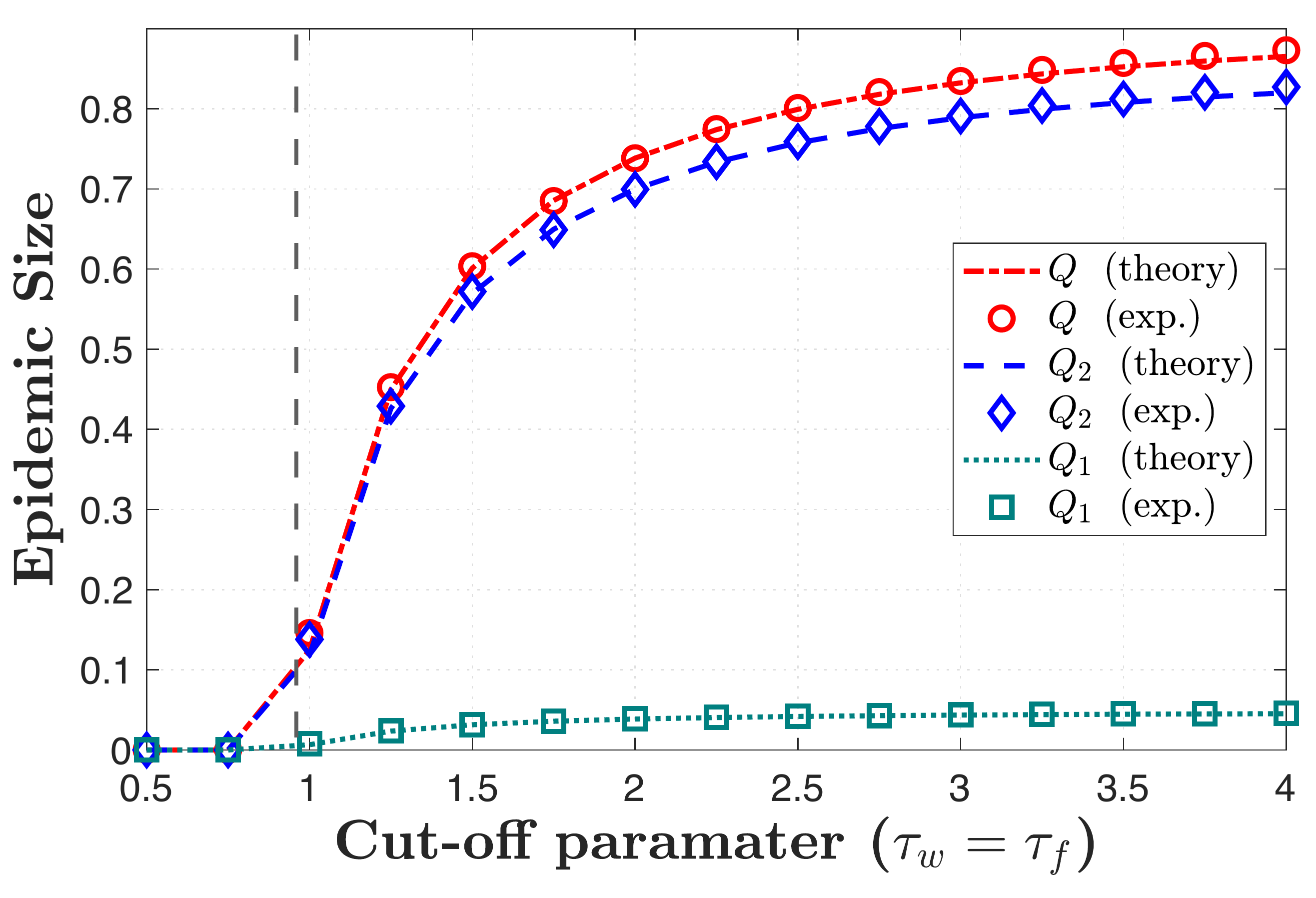}
		\caption{\sl The probability of emergence and mean epidemic size for network layers following power-law degree distribution with exponential cut-off parameters $\tau_f$ and $\tau_w$, averaged over 500 independent experiments.} 
		\label{fig:ne:plot2}
	\end{figure}

Next, we study power law degree distributions with gathering limits modeled through exponential cutoffs parameterized as follows. %, the power law degree distribution can be
	\begin{align}
	&d^f =	\begin{cases}
		0 ~~\text{for } k=0
		\\
		{k^{- \nu_f}}{e^{-{k}/{\tau_f}}}\left({\rm Li}_{\nu_f}({e^{-{1}/{\tau_f}}})\right)^{-1}~~~~\text{for } k \geq 1
	\end{cases}
	\nonumber\\& d^w =	\begin{cases}
		0 ~~\text{for } k=0
		\\
		{k^{- \nu_w}}{e^{-{k}/{\tau_w}}}\left({\rm Li}_{\nu_w}({e^{-{1}/{\tau_w}}})\right)^{-1}~~~~\text{for } k \geq 1,
	\end{cases}
\end{align}
where ${\rm Li}_{\nu}(z)=\sum_{k=1}^{\infty} k^{-\nu}z^k$ is the polylogarithm of order $\nu$ with argument $z$. In Figure~\ref{fig:ne:plot2}, we plot the probability of emergence and mean epidemic size averaged over 500 independent experiments with the vertical dashed line indicating the epidemic threshold $\rho(\pmb{J})=1$. To highlight the impact of varying the cutoff parameter, we set $\tau_f=\tau_w$ and vary it in the interval $[1,5]$, while keeping the other parameters fixed at
$\tfone =0.65, \tftwo = 0.7 $, $\twone=0.3, \twtwo=0.4, \mufoneone=\muwoneone=0.1 , \muftwotwo=\muwtwotwo=0.95$, $\nu_f=2.5$, and $\nu_w=2.3 $. We again notice that the analytical predictions in Theorems 1-3 align well with the simulations.
\section{Additional Results under One-step Irreversible Mutations}
The analysis in this section focuses on one-step irreversible \cite{ne:alexday,antia2003role} where a simple change is required for the contagion to evolve to a highly transmissible variant. Throughout, we assume $\pmb{\mu}=\pmb{\mu}^f=\pmb{\mu}^w$) and the following structure for mutation and transmission matrices:
\begin{align}
		\pmb{\mu}= \left[   \begin{matrix}
			\mu_{11} &  1-\mu_{11}\\ 
			0 &  1
		\end{matrix} \right], 0<\muoneone<1 %\label{eq:onesteprev}
	\end{align}
		\begin{align}
	\left[   \begin{matrix}
		\twone &  0\\ 
		0 &  \twtwo
	\end{matrix} \right]= c\left[   \begin{matrix}
		\tfone & 0\\ 
		0 &  \tftwo
	\end{matrix} \right], c\geq1;~~ T^w_1<T^w_2 {\rm{~and~}} \tfone<T^f_2% \label{eq:transmission_coupling}.
\end{align}
Recall from Lemma 1 that with ${\twone}/{\tfone} ={\twtwo}/{\tftwo}=c$, where $c>0$, and  $\pmb{\mu}=\pmb{\mu}^f=\pmb{\mu}^w$, it holds that
	\begin{align}
		\sigma(\pmb{J}) = \sigma \left(	\left[   \begin{matrix}
			\beta_f &  c \lambda_w\\\ 
			\lambda_f &  c \beta_w
		\end{matrix} \right] \right) \times \sigma(\pmb{T^f} \pmb{\mu}).% \label{eq:spec-prod}
	\end{align}
\subsection{Epidemic Threshold} We first derive the epidemic threshold for the case of one-step irreversible mutations.\\
Our goal is to prove Lemma 2 which states
\begin{align}
		\sigma(\pmb{J}) = \tftwo \times \sigma \left(	\left[   \begin{matrix}
			\beta_f &  c \lambda_w\\\ 
			\lambda_f &  c \beta_w
		\end{matrix} \right] \right) . %\label{eq:spec-prod-irrev}
	\end{align}
\emph{Proof of Lemma 2:}
 Note that $\pmb{\mu}$ is one-step irreversible, and the matrix $\pmb{T^f} \pmb{\mu}$ is upper triangular. Thus,  $ \sigma(\pmb{T^f} \pmb{\mu})=\max\{\tfone \muoneone, \tftwo\}$. Since strain-2 is more transmissible ($\tfone < \tftwo$), we have $ \sigma(\pmb{T^f} \pmb{\mu})= \tftwo$ and \eqref{eq:spec-prod} implies that $\sigma(\pmb{J})$ is not impacted by the magnitude of $\muoneone$ or $\tfone$.%)% (same argument should probably work even when  $\mu_{22}=1-\epsilon$ as long as  $T_1 \muoneone < T_2 (1-\epsilon)$ ) 
	\hfill $\blacksquare$
\subsection{Probability of Mutation}
\begin{figure}[t]
\centering
	\includegraphics[scale=0.44]{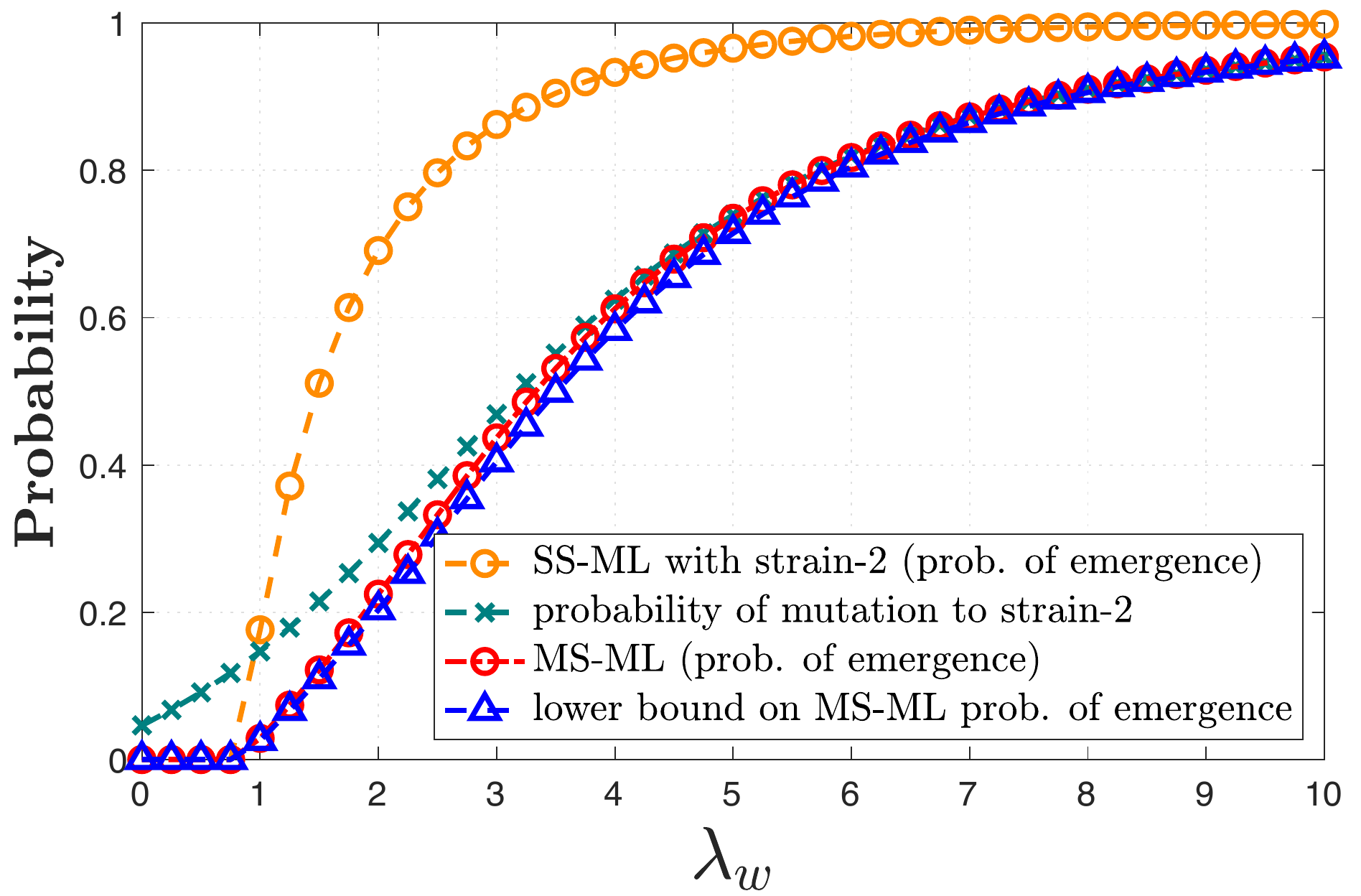}
\caption{\sl A lower bound for the probability of emergence when the mutation matrix is one step irreversible with $\mutwotwo \rightarrow 1$. The probability of emergence for the multi-layer multi-strain model can be expressed as a product of two factors- the probability of mutation to strain-2 along the chain of transmissions and the probability of emergence if only strain-2 was circulating in the population. Here, the network layers are taken to be Poisson with parameters $\lambda_f=1$ and $\lambda_w \in [0,10]$. } 
\label{fig:ne:lb-prob}
\end{figure}

Next,  we derive an interpretable bound for probability of emergence by evaluating the probability of mutation, following the approach in \cite{ne:pnas}. Throughout this discussion, we assume that \eqref{eq:onesteprev} holds and $\tfone<\tftwo, \twone<\twtwo$.
	We have	
	\begin{align}
	&\pr[\rm{emergence}]\geq\pr[\rm{emergence}~|~\rm{~at~least~ 1~ mutation~to~strain-2}~] \cdot\pr[ \rm{at~least~1~mutation~to~strain-2}~]\label{eq:ne-bound1}.
% 	&\quad+\pr[\rm{emergence}~|~\rm{no~ mutation~to~strain-2}~] \nonumber\\
% 	&\quad\cdot(1-\pr[ \rm{at~least~one~mutation~to~strain-2}~])
	\end{align}
Let $\mathcal{E}_1$ denote the event that there is a positive fraction of strain-1 infections before a mutation to strain-2 appears. Further, let $P^{\rm ML-2}$ denote the probability of emergence on the multi-layer network $\mathbb{H}$ when only strain-2 circulates in the population. We have
		\begin{align}
			&\pr[\rm{emergence}~|~\rm{~at~least~ 1~ mutation~to~strain-2}~]\nonumber\\
			%&=
			%\pr[\rm{emergence}~|~\rm{~at~least~ 1~ mutation~to~strain-2},~ \mathcal{E}_1] \pr[\mathcal{E}_1]+\pr[\rm{emergence}~|~\rm{~at~least~ 1~ mutation~to~strain-2},~ \mathcal{E}_1^c ] \pr[\mathcal{E}_1^c]
						&=
			\pr[\rm{emergence},~ \mathcal{E}_1~|~\rm{~at~least~ 1~ mutation~to~strain-2}]+\pr[\rm{emergence},~ \mathcal{E}_1^c ~|~\rm{~at~least~ 1~ mutation~to~strain-2}] \nonumber
			\\&\geq 			 \pr[\rm{emergence},~ \mathcal{E}_1^c ~|~\rm{~at~least~ 1~ mutation~to~strain-2}] \label{eq:ne-bound2-penult}\\
			&= P^{\rm ML-2}\label{eq:ne-bound2}.
		\end{align}
%\end{enumerate}
Note that the bound in \eqref{eq:ne-bound2-penult} is tight, when the transmissibility of strain-1 is below the critical threshold, i.e., when 
\begin{align}
\tfone  \rho \left(	\left[   \begin{matrix}
		\beta_f &  c \lambda_w\\\ 
		\lambda_f &  c \beta_w
	\end{matrix} \right]\right) \leq1. \label{eq:contrast-2}
\end{align}
Combining \eqref{eq:ne-bound1} and \eqref{eq:ne-bound2}, we get
	\begin{align}
	\pr[\rm{emergence}]&\geq P_2^{\rm ML} \pr[ \rm{at~least~1~mutation~to~strain-2}~] \label{eq:ne-lowerbprob}.
\end{align}
%(Aside: haven't used the specific form of distributions till this point)]]\\
In the above lower bound, $P_2^{\rm ML}$ can be obtained from Theorem~1 by substituting $(\tfone,\twone) \leftarrow (\tftwo, \twtwo)$. For computing $\pr[ \rm{at~least~1~mutation~to~strain-2}~]$, we find the complementary probability, i.e., the $\pr[ \rm{no~mutation~to~strain-2}~]$
	%	\emph{Impact of mutation matrix on threshold- think about the impossibility for result obtained while trying to contrast two-strain/two-layer vs single-strain/two-laye
	by solving a system of recursive equations. We first obtain $\pr[ \rm{no~mutation~to~strain-2}~]$ in the chain of infection events emanating from a later generation infective reached by type-$f$ edge (resp., type-w edge), which in turn yields the probability of no mutation starting from the initial infective (seed node). For, $a\in \{f,w\}$, let $q^a$ denote the probability of there being no mutation to strain-2 in the chain of infections emanating from a later generation infective that was infected through a type-$a$ edge. We have
%	{\color{red}Add proof}
%	\emph{Conjecture} (Needs proof and exp. verification)\\
\begin{align}
  	&\pr[ \rm{at~least~1~mutation~to~strain-2}~]\nonumber\\ &=1-g(1-T^f_1+T^f_1\mu_{11}q^f,1-T^w_1+T^w_1\mu_{11}q^w),  %\label{ne:eq:bound-rec-0}
\end{align}
where,
	\begin{align}
		q^f &= G^f(1-T^f_1+T^f_1\mu_{11}q^f,1-T^w_1+T^w_1\mu_{11}q^w),\\%\label{ne:eq:bound-rec-1}
		q^w &= G^w(1-T^f_1+T^f_1\mu_{11}q^f,1-T^w_1+T^w_1\mu_{11}q^w),%\label{ne:eq:bound-rec-2}
	\end{align}
% and the PGFs $g, G^f, G^w$ are respectively defined through \eqref{eq:small-g}-\eqref{eq:big-g-w}.
To see why \eqref{ne:eq:bound-rec-0} holds, note that for no mutation to strain-2 to occur, each susceptible neighbor of a later-generation must either get infected with strain-1 or remain uninfected. Let $\bar{p}_{d^f}$ (respectively, $\bar{p}_{d^w}$) denote the excess degree distribution of a later generation infective reached by following a type-$f$ (respectively, type-$w$ edge).
	\begin{align}
q^f&= \sum_{d^f}\sum_{d^w} \Bigg(\bar{p}_{d^f} \bar{p}_{d^w}\sum_{k^f=0}^{d^f}(1-\tfone)^{d^f-k^f}(\tfone \mu_{11}q^f)^{k^f}\nonumber\\
 &~~\cdot \sum_{k^w=0}^{d^w}(1-\twone)^{d^w-k^w}(\twone \mu_{11}q^w)^{k^w}\Bigg) \nonumber\\
& =\sum_{d^f}\sum_{d^w} \bar{p}_{d^f} \bar{p}_{d^w}(1-\tfone+\tfone\mu_{11}q^f)^{d^f}\nonumber\\
&\quad\cdot (1-\twone+\twone\mu_{11}q^w)^{d^w}\nonumber\\
&=G^f(1-T^f_1+T^f_1\mu_{11}q^f,1-T^w_1+T^w_1\mu_{11}q^w).\nonumber
	\end{align}
In Figure~\ref{fig:ne:lb-prob}, we plot the lower bound obtained through \eqref{eq:ne-lowerbprob} and compare it with the probability of emergence obtained through Theorem~1. The degree distribution for layers $f$ and $w$ follow Poisson distributions with parameters $\lambda_f$ and $\lambda_w$ respectively. We set $\tfone=0.2$, $\tftwo =0.5$, $\twone=0.3$, $\twtwo=0.6$; $\mufoneone=\muwoneone=0.8$; $\lambda_f=1$, and vary $\lambda_w$. We observe a tight correspondence between the lower bound and the probability of emergence. We note that as more edges are added to layer-$w$, the more likely it is for a mutation to the highly transmissible strain-2, which ultimately makes the outbreak more likely.
\section{Reduction to Simpler Models}
\subsection{The case of Poisson network layers}
Suppose, the degree distribution for the network-layers $\mathbb{F}$ and $\mathbb{W}$ follow the distribution Poisson($\lambda_f$) and  Poisson($\lambda_w$) respectively.
Now consider the following transformations that reduce the two network layers into a single Poisson network layer with mean degree $\lambda$ given by the sum of the mean degrees of the constituent layers. For each strain $i=1,2$, we take $T_i$ to be the the weighted average of the transmissibilities in the two layers where the weights correspond to the mean degree in each layer. This gives
\begin{align}
\lambda &\leftarrow \lambda_f +  \lambda_w.\\% \label{eq:mapto1layerlambda}
T_1 &\leftarrow \frac{\lambda_f  \tfone+ \lambda_w \twone}{\lambda_f + \lambda_w};~~ T_2 \leftarrow \frac{\lambda_f  \tftwo+ \lambda_w \twtwo}{\lambda_f + \lambda_w}. %\label{eq:mapto1layer} 
	\end{align}

	Next, we show that when both network layers are purely Poisson, the probability of emergence predicted by the multi-strain model on a single through the mapping \eqref{eq:mapto1layerlambda} and \eqref{eq:mapto1layer} is accurate. When a network layer is Poisson, say with parameter $\lambda$, the PGF for the degree distribution (denoted, $g(z)$) and excess degree distribution (denoted, $G(z)$) is given as $g(z)=G(z)=e^{-\lambda(1-z)}$. To prove the above, we note that when both layers are independent Poisson distributions, the analytical probability of emergence for the multi-layer model is given as follows. 
\begin{align}
			\pr[\rm{emergence}]&=\gamma_1(q^f_1,q^f_2, q^w_1,q^w_2 )\nonumber\\
			&= g \Bigg(1-\tfone+\tfone \left( \sum_{j=1}^2 \mu_{1j} q^f_j\right), 1-\twone+\twone \left(\sum_{j=1}^2 \mu_{1j} q^w_j\right) \Bigg),\nonumber\\
			&= \exp\left\{\lambda_f\left(-\tfone+\tfone \left( \sum_{j=1}^2 \mu_{1j} q^f_j\right)\right)\right\}\cdot\exp\left\{\lambda_w\left(-\twone+\twone \left( \sum_{j=1}^2 \mu_{1j} q^w_j\right)\right)\right\} \label{eq:map_to_one_layer_poisson_q},
		\end{align}
		where for $a \in \{f,w\}$ and $i \in \{1,2 \}$,
		\begin{align}
			q^a_i &=  \Gamma^a_i(q^f_1,q^f_2, q^w_1,q^w_2 )\nonumber\\ 
			&= G^a \Bigg(1-T^f_i+T^f_i \left(\sum_{j=1}^2 \mu^f_{ij} z^f_j \right),1-T^w_i+T^w_i \left(\sum_{j=1}^2 \mu^w_{ij} z^w_j \right)\Bigg),\nonumber\\
			&= \exp\left\{\lambda_f\left(-\tfone+\tfone \left( \sum_{j=1}^2 \mu_{1j} q^f_j\right)\right)\right\}\cdot\exp\left\{\lambda_w\left(-\twone+\twone \left( \sum_{j=1}^2 \mu_{1j} q^w_j\right)\right)\right\}.\nonumber
			%&= \exp\left\{-(\lambda_f +\lambda_w)\right\}
		\end{align}
		where the last step follows from the independence of the degree distribution of the network layers and the fact that for the Poisson degree distribution, the PGF of the excess degree distribution shares the same functional form as the PGF of the degree distribution. Using \eqref{eq:map_to_one_layer_poisson_q}, we have $\qfone=\qwone$ and $\qftwo = \qwtwo$. Substituting in \eqref{eq:map_to_one_layer_poisson_q},
		\begin{align}
		&	\pr[\rm{emergence}]\nonumber\\
		&= \exp\left\{-(\lambda_f+\lambda_w)\frac{\lambda_f  \tfone+ \lambda_w \twone}{\lambda_f + \lambda_w} \left(-1+\left( \sum_{j=1}^2 \mu_{1j} q^f_j\right)\right)\right\}. 
		\end{align}

\begin{figure}[t]
\centering
	\includegraphics[scale=0.42]{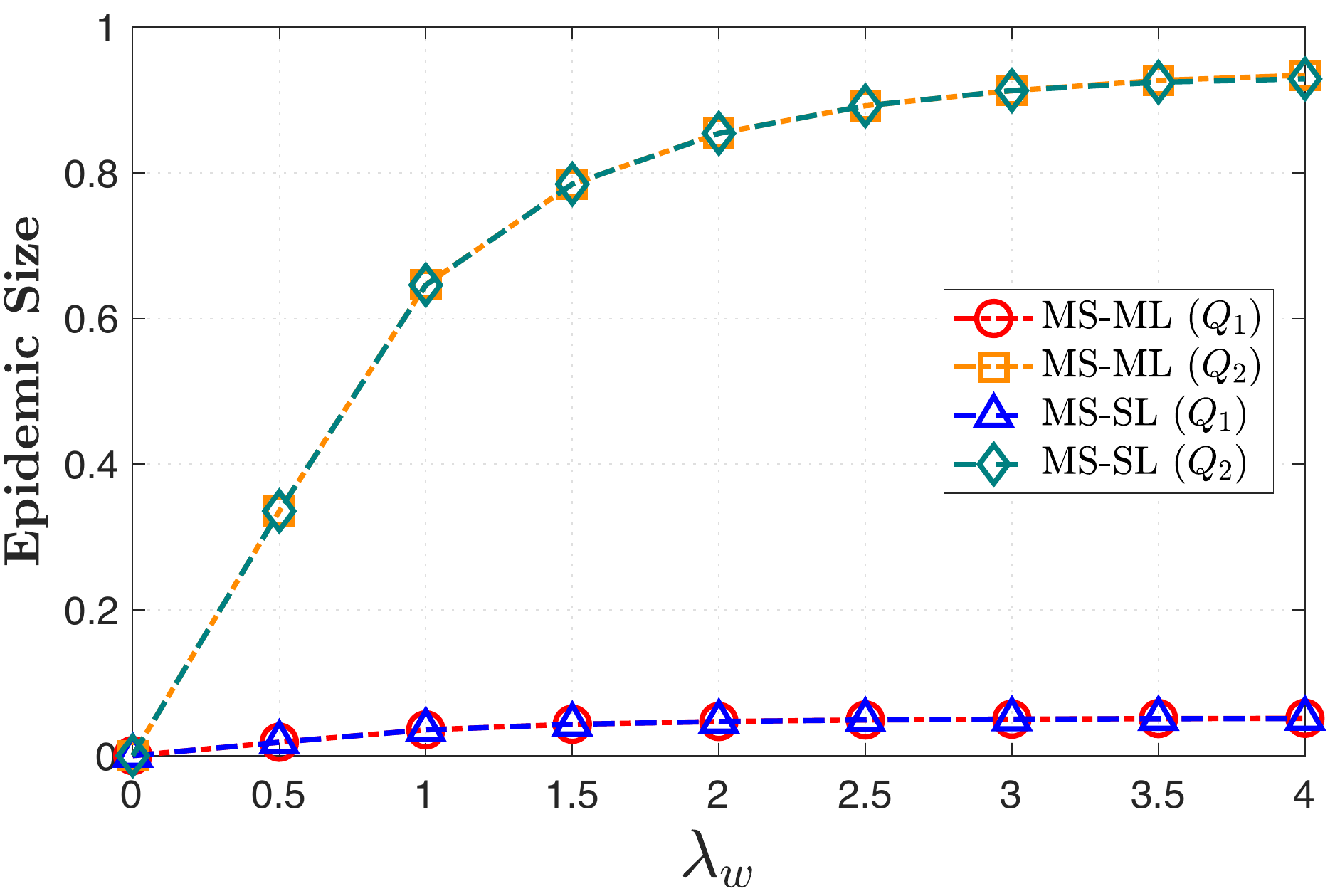}
\caption{\sl For multi-strain spreading on a network comprising of two independent Poisson layers with parameters $\lambda_f=1$ and $\lambda_w$, we plot the epidemic size obtained through Theorem 3 indicated as MS-ML. We also plot the corresponding prediction made by reduction to a single-layer\cite{ne:pnas} through the transformations \eqref{eq:mapto1layer} and \eqref{eq:mapto1layerlambda}, indicated as MS-SL.} 
\label{fig:ne:pure-poisson}
\end{figure}

		This is in line with the prediction on a single-layer model \cite{ne:pnas,ne:alexday} through the mapping \eqref{eq:mapto1layer} and \eqref{eq:mapto1layerlambda}. Since, we obtain the same system of recursive equations for the probability of emergence for the multi-layer model as with the reduction to the  single-layer model, the corresponding epidemic threshold is also the same. Next, we plot the analytical epidemic size as predicted by reduction to a single-layer model through \eqref{eq:mapto1layerlambda} and \eqref{eq:mapto1layer}, when in fact there are two distinct layers in the network; see Figure~\ref{fig:ne:pure-poisson}. We set	$\tfone=0.6,\tftwo=0.8, \twone=0.7, \twtwo=0.9,\muoneone=0.1, \mutwotwo=0.95, \lambda_f=1 $, and vary $\lambda_w$ in the interval $[0,4]$. We observe that the fraction of infected individuals for each type of strain ($Q_1$ and $Q_2$) is accurately predicted by the multi-strain model on a single Poisson layer with mean degree $\lambda_f+\lambda_w$.

%'appears' similar empirically (experiments- approximation by clipping sum).
	%	{{\color{red}[Add]}- plot} for 
% 	for pure Poisson

\subsection*{The impact of dispersion indices}

Recall that under the transformation to a single-layer network obtained by taking the sum of node degrees, the epidemic threshold is predicted correctly
when
\begin{align}
				\sigma \left( \left[   \begin{matrix}
		\beta_f &  \lambda_w\\\ 
		\lambda_f &   \beta_w
	\end{matrix} \right] \right) =  \frac{\beta_f \lambda_f + \beta_w \lambda_w +2 \lambda_f \lambda_w}{\lambda_f+\lambda_w}  %\label{eq:pred-thresh-1eq2}.
\end{align}
It can be verified that with $\lambda_f, \lambda_w, \beta_f, \beta_w>0$,  \eqref{eq:pred-thresh-1eq2} holds if and only if 
\begin{align}
	\beta_f - \lambda_f =\beta_w -\lambda_w %\label{eq:matching-condition}.
\end{align}

Next, we prove that \eqref{eq:matching-condition} is necessary and sufficient for \eqref{eq:pred-thresh-1eq2} to hold.
Namely, implying that matching the epidemic threshold through a reduction to a single layer with degree distribution (given as the sum of the degree distribution of the constituent layers) requires the dispersion index of constituent layers to be equal. Therefore, our goal is to show that
\begin{align}
				\sigma \left( \left[   \begin{matrix}
		\beta_f &  \lambda_w\\\ 
		\lambda_f &   \beta_w
	\end{matrix} \right] \right) =  \frac{\beta_f \lambda_f + \beta_w \lambda_w +2 \lambda_f \lambda_w}{\lambda_f+\lambda_w} \iff \beta_f - \lambda_f =\beta_w -\lambda_w \label{eq:iff-quadratic}.
\end{align}\emph{Proof:}
We have,
\begin{align}
\sigma \left( \left[   \begin{matrix}
		\beta_f &  \lambda_w\\\ 
		\lambda_f &   \beta_w
	\end{matrix} \right] \right) =  \frac{\beta_f +\beta_w + \sqrt{(\beta_f-\beta_w)^2+4\lambda_f \lambda_w}}{2} \label{eq:spec-expand}
\end{align}
Substituting in \eqref{eq:iff-quadratic} and rearranging, we see that the proof of \eqref{eq:iff-quadratic} requires 
\begin{align}
   & \sqrt{(\beta_f-\beta_w)^2+4\lambda_f \lambda_w} =  2 \times\frac{\beta_f \lambda_f + \beta_w \lambda_w +2 \lambda_f \lambda_w}{\lambda_f+\lambda_w}  - (\beta_f + \beta_w)\nonumber.\\
\end{align}
Further, note that 
\begin{align}
   2\times  \frac{\beta_f \lambda_f + \beta_w \lambda_w +2 \lambda_f \lambda_w}{\lambda_f+\lambda_w}  - (\beta_f + \beta_w) %&= \frac{2(\beta_f \lambda_f + \beta_w \lambda_w +2 \lambda_f \lambda_w)- (\beta_f+\beta_w)(\lambda_f+\lambda_w)}{{\lambda_f+\lambda_w}}\nonumber\\
   % &= \frac{\beta_f \lambda_f + \beta_w \lambda_w+4 \lambda_f\lambda_w - \beta_f \lambda_w - \beta_w \lambda_f}{\lambda_f+\lambda_w} \nonumber\\
     &= \frac{(\beta_f-\beta_w)(\lambda_f-\lambda_w)+ 4 \lambda_f\lambda_w }{\lambda_f+\lambda_w}. \nonumber
\end{align}
Therefore, we need to show that
\begin{align}
   & {(\beta_f-\beta_w)^2+4\lambda_f \lambda_w} = \left[\frac{(\beta_f-\beta_w)(\lambda_f-\lambda_w)+ 4 \lambda_f\lambda_w }{\lambda_f+\lambda_w}\right]^2 \label{eq:disp-vs2}.
\end{align}
Substituting $\beta_f-\beta_w \rightarrow t$ in \eqref{eq:disp-vs2} and rearranging, we need to show that
\begin{align}
 \left[\frac{t(\lambda_f-\lambda_w)+ 4 \lambda_f\lambda_w }{\lambda_f+\lambda_w}\right]^2 - t^2 = 4 \lambda_f \lambda_w,
\end{align}
or equivalently
\begin{align}
   % \left[ \frac{2\lambda_f}{\lambda_f+\lambda_w}t +\frac{4 \lambda_f\lambda_w }{\lambda_f+\lambda_w}\right]\left[ \frac{-2\lambda_w}{\lambda_f+\lambda_w}t +\frac{4 \lambda_f\lambda_w }{\lambda_f+\lambda_w}\right]=4 \lambda_f \lambda_w,\nonumber\\
        \left( t +{2\lambda_w }\right)\left( {-}t +{2 \lambda_f }\right)= ({\lambda_f+\lambda_w})^2. \label{eq:disp-intermed}
\end{align}
From \eqref{eq:disp-intermed}, it is evident that the line $f_1(t)=({\lambda_f+\lambda_w})^2$ is a tangent to the parabola $f_2(t)=\left( t +{2\lambda_w }\right)\left( {-}t +{2 \lambda_f }\right)$ at $t=\lambda_f-\lambda_w$, and thus, $t=\beta_f-\beta_w = \lambda_f-\lambda_w$ emerges as the unique solution to \eqref{eq:disp-intermed}; consequently \eqref{eq:iff-quadratic} holds.
\hfill{$\blacksquare$}\\

% Comparing the above thresholds in (\ref{eq:pred-thresh-2}) and (\ref{eq:pred-thresh-1}), we see that the predicted thresholds are identical if and only if
% \begin{align}
% 				\sigma \left( \left[   \begin{matrix}
% 		\beta_f &  \lambda_w\\\ 
% 		\lambda_f &   \beta_w
% 	\end{matrix} \right] \right) =  \frac{\beta_f \lambda_f + \beta_w \lambda_w +2 \lambda_f \lambda_w}{\lambda_f+\lambda_w}  \label{eq:pred-thresh-1eq2}.
% \end{align}
% It can be verified that with $\lambda_f, \lambda_w, \beta_f, \beta_w>0$,  \eqref{eq:pred-thresh-1eq2} holds if and only if 
% \begin{align}
% 	\beta_f - \lambda_f =\beta_w -\lambda_w \label{eq:matching-condition}.
% \end{align}
% {\color{red} Type proof}

\section{Further Reductions and Challenges}
\label{sec:further-comparisons}
 Next, we propose and evaluate different transformations for finding a corresponding single-strain multi-layer (SS-ML) or a multi-strain single-layer (MS-SL) model for a given MS-ML model.  In what follows, it will be useful to define the following quantity: for $i \in \{1,2 \}$ and $a\in \{f, w\}$, we define $\overline{T}_i^a$ to be the probability that a given vertex is infected with strain $i$ through layer $a$. Mathematically, this can be written as 
\begin{equation}
\overline{T}_i^a : = T_1^a\mu_{1i} + T_2^a \mu_{2i}. 
\end{equation}
\subsection*{Additional challenges with transformations to a single layer (MS-SL)} 
We have seen challenges associated with reduction to MS-SL models by taking the sum of the degrees in the two layers as the degree for an equivalent single layer. Here, we outline another method for transforming to a single layer by matching the \emph{mean matrix}; that is, we ensure that the mean number of secondary infections stemming from any given type of infected individual is the same across the models. For the branching process corresponding to the MS-SL model, we say that an infected vertex is type 1 if it has been infected with strain 1 and type 2 otherwise. The mean matrix of this model is given by 
\begin{equation}
\label{eq:M_ms-sl}
\mathbf{M} : = \beta \begin{pmatrix}
T_1 \mu_{11} & T_1 \mu_{12} \\
T_2 \mu_{21} & T_2 \mu_{22}
\end{pmatrix},
\end{equation}
where, the $(i,j)$ entry of the matrix denotes the expected number of type-$j$ secondary infections caused by a type-$i$ infective. \\Let $\mathbf{J}^{MS-SL}$ denote the version of $\mathbf{J}$ corresponding to the MS-SL model. By carrying out similar arguments as for the SS-ML reduction and noting that 
\begin{align*}
\mathbb{P} ( \text{node infected through layer $f$} \vert \text{has strain 1} ) & = \frac{ \overline{T}_1^f }{ \overline{T}_1^f + \overline{T}_1^w} \\
\mathbb{P} ( \text{node infected through layer $w$} \vert \text{has strain 2}) & = \frac{ \overline{T}_1^w }{ \overline{T}_1^f + \overline{T}_1^w },
\end{align*}
we can then write
\begin{align*}
\mathbf{J}_{11}^{MS-SL} &= \left( \frac{ \overline{T}_1^f }{ \overline{T}_1^f + \overline{T}_1^w} ( T_1^f \beta_f + T_1^w \lambda_w) \right. \left. +  \frac{ \overline{T}_1^w }{ \overline{T}_1^f + \overline{T}_1^w} (T_1^w \beta_w + T_1^f \lambda_f)   \right) \mu_{11}.
\end{align*}
More generally, we have for $i,j \in \{1,2 \}$ that 
\begin{align}
\mathbf{J}_{ij}^{MS-SL} & = \left( \frac{ \overline{T}_i^f }{ \overline{T}_i^f + \overline{T}_i^w} ( T_i^f \beta_f + T_i^w \lambda_w) \right. 
\label{eq:J_ms-sl}
\left. +  \frac{ \overline{T}_i^w }{ \overline{T}_i^f + \overline{T}_i^w} (T_i^w \beta_w + T_i^f \lambda_f)   \right) \mu_{ij}.
\end{align}
From the general formula in \eqref{eq:J_ms-sl}, it is not straightforward to map the MS-ML model to the MS-SL model since the network parameters are intertwined with the transmissibilities in \eqref{eq:J_ms-sl}, whereas the network parameters are clearly separable from the viral transmission properties in \eqref{eq:M_ms-sl}. However, in the special case where $T_1^w / T_1^f = T_2^w / T_2^f = c$, some simplifications can be made. Under this assumption, it holds that\begin{align} \overline{T}_i^f / ( \overline{T}_i^f + \overline{T}_i^w ) = 1/ (1 + c),\nonumber
\end{align} 
hence,
\begin{align}
\mathbf{J}_{ij}^{MS-SL} & = \left( \frac{1}{1 + c} ( \beta_f + c \lambda_w) + \frac{ c^2 }{1 + c} (c \beta_w + \lambda_f) \right) T_i^f \mu_{ij} \nonumber\\
& = : \widetilde{\beta} T_i^f \mu_{ij}.\nonumber
\end{align}
This indicates that a reasonable way to set the parameters of a corresponding MS-SL model with a matching mean matrix is to use the transmissibilities of layer $f$ along with the \emph{effective} mean excess degree parameter $\widetilde{\beta}$. However, such a transformation does not provide a systematic way to infer the exact probability distribution for the single-layer, which is critical to predicting the epidemic characteristics using a MS-SL model.  

\subsection{Reductions to single-strain multi-layer (SS-ML) models}
Next, we more concretely evaluate various reductions that can be made to translate a MS-ML model to a SS-ML model. For reductions to SS-ML models, assuming $\frac{\twone}{\tfone} =\frac{\twtwo}{\tftwo}=c$, we consider the following two approaches. The first approach involves using \eqref{eq:spec-prod} and defining the equivalent transmissibilities for the two layers as:
\begin{align}
    &\rho(\pmb{T^f}\pmb{\mu}) \rightarrow \tilde{T}^f, \nonumber\\
        &\rho(\pmb{T^w}\pmb{\mu}) \rightarrow \tilde{T}^w (\equiv c \rho(\pmb{T^f}\pmb{\mu}) \rightarrow \Tilde{T}^w). %\label{eq:ss-tx-rho}
\end{align}
Through \eqref{eq:spec-prod}, the transformation \eqref{eq:ss-tx-rho} ensures that the spectral radius predicted by the corresponding SS-ML reduction is the same as the MS-ML model. 
The second approach is based on matching the mean matrix as done for the MS-SL case. 
% \hrule
% 	\begin{itemize}
% 		\item (Mapping the epidemic threshold)%spectral radius) 
% % $(p^f, \lambda^f \beta_f), (p^w, \lambda^w \beta_w)$

% Using \eqref{eq:spec-prod}
% $$
% \mathbf{J}^{SS-ML} : =\begin{pmatrix}
% 	 \tilde{T}_f 	\beta_f &    \tilde{T}_w  \lambda_w\\\
% 	 \tilde{T}_f		\lambda_f &  \tilde{T}_w \beta_w
% \end{pmatrix}
% $$
% \hrule
% \item (Mapping the Mean Matrix) \\
% %{\color{green}To Add}
% \begin{align}
% &\dfrac{\tfone \bar{T}^f_1  + \tftwo \bar{T}^f_2}{\tfone+\tftwo} \rightarrow \tilde{T}^f \nonumber \\
% &\dfrac{\twone \bar{T}^w_1  + \twtwo \bar{T}^w_2}{\twone+\twtwo} \rightarrow \tilde{T}^w, \nonumber \\
% ( \equiv~ &c \cdot \dfrac{\tfone \bar{T}^f_1  + \tftwo \bar{T}^f_2}{\tfone+\tftwo}\rightarrow \tilde{T}^w )
% \end{align}
% where, 
% \begin{align}
% &\bar{T}^f_1 = \tfone \mu_{11}+\tftwo \mu_{21} \nonumber\\
% &\bar{T}^f_2 = \tfone \mu_{12}+\tftwo \mu_{22}
% \end{align}
% \end{itemize}
% $$
% \mathbf{J}^{SS-ML} : = \begin{pmatrix}
% \tilde{T}_f \beta_f  & \tilde{T}_w \lambda_w \\
% \tilde{T}_f \lambda_f  & \tilde{T}_w \beta_w
% \end{pmatrix}
% $$
In the branching process corresponding to the SS-ML model, there are two types of infected vertices: those that have been infected through layer $f$ (type 1) and those that have been infected through layer $w$ (type 2). The mean matrix for this model is given by
$$
\mathbf{M} : = \begin{pmatrix}
T^f \beta_f  & T^w \lambda_w \\
T^f \lambda_f  & T^w \beta_w
\end{pmatrix},
$$
where the $(i,j)$ entry represents the expected number of type-$j$ secondary infectives caused by a type-$i$ infective. Let $\mathbf{J}^{SS-ML}$ denote the version of $\mathbf{J}$ that corresponds to the SS-ML model. To compute the $(i,j)$ entry of $\mathbf{J}^{SS-ML}$, we take an average over the probabilities that a given type-$i$ node with a particular strain infects a type-$j$ node. For instance, noting that 
\begin{align*}
%\label{eq:prob_1f}
\mathbb{P} ( \text{node has strain 1} \vert \text{infected through layer $f$} ) & = \frac{ \overline{T}_1^f }{\overline{T}_1^f + \overline{T}_2^f } \\
%\label{eq:prob_2f}
\mathbb{P} ( \text{node has strain 2} \vert \text{infected through layer $f$} ) & = \frac{ \overline{T}_2^f }{\overline{T}_1^f + \overline{T}_2^f },
\end{align*}
we can then write
\begin{align}
\mathbf{J}_{11}^{SS-ML} = \left( \frac{ \overline{T}_1^f }{ \overline{T}_1^f + \overline{T}_2^f}  T_1^f  + \frac{ \overline{T}_2^f}{ \overline{T}_1^f + \overline{T}_2^f } T_2^f \right) \beta_f . \nonumber
\end{align}
Similarly, we have that 
\begin{align*}
\mathbf{J}_{12}^{SS-ML} & = \left( \frac{ \overline{T}_1^f }{ \overline{T}_1^f + \overline{T}_2^f}  T_1^w  + \frac{ \overline{T}_2^f}{ \overline{T}_1^f + \overline{T}_2^f } T_2^w \right) \lambda_w \end{align*}\begin{align*}
\mathbf{J}_{21}^{SS-ML} & = \left( \frac{ \overline{T}_1^w }{ \overline{T}_1^w + \overline{T}_2^w}  T_1^f  + \frac{ \overline{T}_2^w}{ \overline{T}_1^w + \overline{T}_2^w} T_2^f \right) \lambda_f \\
\mathbf{J}_{22}^{SS-ML} & = \left( \frac{ \overline{T}_1^w }{ \overline{T}_1^w + \overline{T}_2^w}  T_1^w  + \frac{ \overline{T}_2^w}{ \overline{T}_1^w + \overline{T}_2^w} T_2^w \right) \beta_w .
\end{align*}
Note that under the above transformation, the MS-ML model can be mapped to SS-ML in the special case where $T_1^w / T_1^f = T_2^w / T_2^f = c$, it holds that 
$$
\frac{ \overline{T}_1^f }{\overline{T}_1^f + \overline{T}_2^f} = \frac{ \overline{T}_1^w}{ \overline{T}_1^w + \overline{T}_2^w},
$$
hence $\mathbf{J}^{SS-ML}$ is of the same form as $\mathbf{M}$ with 
\begin{align}
T^f & : = \frac{ \overline{T}_1^f }{ \overline{T}_1^f + \overline{T}_2^f}  T_1^f  + \frac{ \overline{T}_2^f}{ \overline{T}_1^f + \overline{T}_2^f } T_2^f, \nonumber\\
T^w & : = \frac{ \overline{T}_1^w }{ \overline{T}_1^w + \overline{T}_2^w}  T_1^w  + \frac{ \overline{T}_2^w}{ \overline{T}_1^w + \overline{T}_2^w} T_2^w = c T^f.  \label{eq:SS-Tx}
\end{align}

\end{document}